\documentclass[aps,pre,twocolumn,superscriptaddress,showpacs,10pt]{revtex4-1}

\usepackage{graphicx}
\usepackage{mathtools}
\usepackage{amssymb,amsmath}
\usepackage[caption = false]{subfig}
\usepackage{booktabs}
\usepackage{placeins}
\usepackage{verbatim}
\usepackage{comment}
\usepackage{color}
\usepackage{float}
\usepackage[utf8]{inputenc}

\renewcommand{\arraystretch}{1.2}
\newcommand{\ra}[1]{\renewcommand{\arraystretch}{#1}}

\newcommand{\etal}[0]{\textit{et al.}\textcolor{white}{a}}
\newcommand{\uv}[1]{{\,\bf \hat{#1}}}

\newcommand{\reffig}[1]{figure~\ref{fig:#1}}
\newcommand{\refeqs}[1]{equation~(\ref{eq:#1})}

\newcommand{\tribra}[1]{\langle{#1}\rangle}
\newcommand{\si}[1]{$\,${#1}}
\newcommand{\rsym}[0]{\mathcal{R}_{xy}}
\definecolor{gray}{rgb}{0.5,0.5,0.5}

\begin{document}

\title{Unstable Equilibria and Invariant Manifolds in Quasi-Two-Dimensional Kolmogorov-like Flow}

\date{\today}

\author{Balachandra Suri}
\email{bsuri@ist.ac.at}
\affiliation{School of Physics, Georgia Institute of Technology, Atlanta, GA 30332, USA}
\affiliation{IST Austria, 3400 Klosterneuburg, Austria}
\author{Jeffrey Tithof}
\affiliation{Department of Mechanical Engineering, University of Rochester, Rochester, NY 14627, USA}
\author{Roman O. Grigoriev}
\affiliation{School of Physics, Georgia Institute of Technology, Atlanta, GA 30332, USA}
\author{Michael F. Schatz}
\affiliation{School of Physics, Georgia Institute of Technology, Atlanta, GA 30332, USA}

\begin{abstract}

Recent studies suggest that unstable, non-chaotic solutions of the Navier-Stokes equation may provide deep insights into fluid turbulence. 
In this article, we present a combined experimental and numerical study exploring the dynamical role of unstable equilibrium solutions and their invariant manifolds in a weakly turbulent, electromagnetically driven, shallow fluid layer. 
Identifying instants when turbulent evolution slows down, we compute 31 unstable equilibria of a realistic two-dimensional model of the flow. 
We establish the dynamical relevance of these unstable equilibria by showing that they are closely visited by the turbulent flow. 
We also establish the dynamical relevance of unstable manifolds by verifying that they are shadowed by turbulent trajectories departing from the neighborhoods of unstable equilibria  over large distances in state space. 

\end{abstract}


\maketitle

\section{Introduction}\label{sec:intro}

Understanding fluid turbulence has remained a long-standing problem in classical physics \cite{hopf_1948, landau_1959, arnold_1964}. 
Recently, substantial progress was made using a theoretical framework that was originally developed \cite{auerbach_1987,cvitanovic_1988} for low-dimensional chaotic systems, such as the Lorentz system \cite{lorenz_1963}.
This framework, which goes back to the work of Poincar\'e on celestial mechanics  \cite{poincare_1993}, uses a hierarchy of unstable temporally simple solutions (e.g., equilibria or periodic orbits) of the governing equations to provide both a dynamical and statistical description of a chaotic system. 
Computing such solutions of the Navier-Stokes equation -- called Exact Coherent Structures (ECSs) -- has become feasible only recently, following the development of novel numerical methods \cite{saad_1986,kelley_1995}.

\begin{figure}[!t]
\begin{center}
{\includegraphics[width=3in]{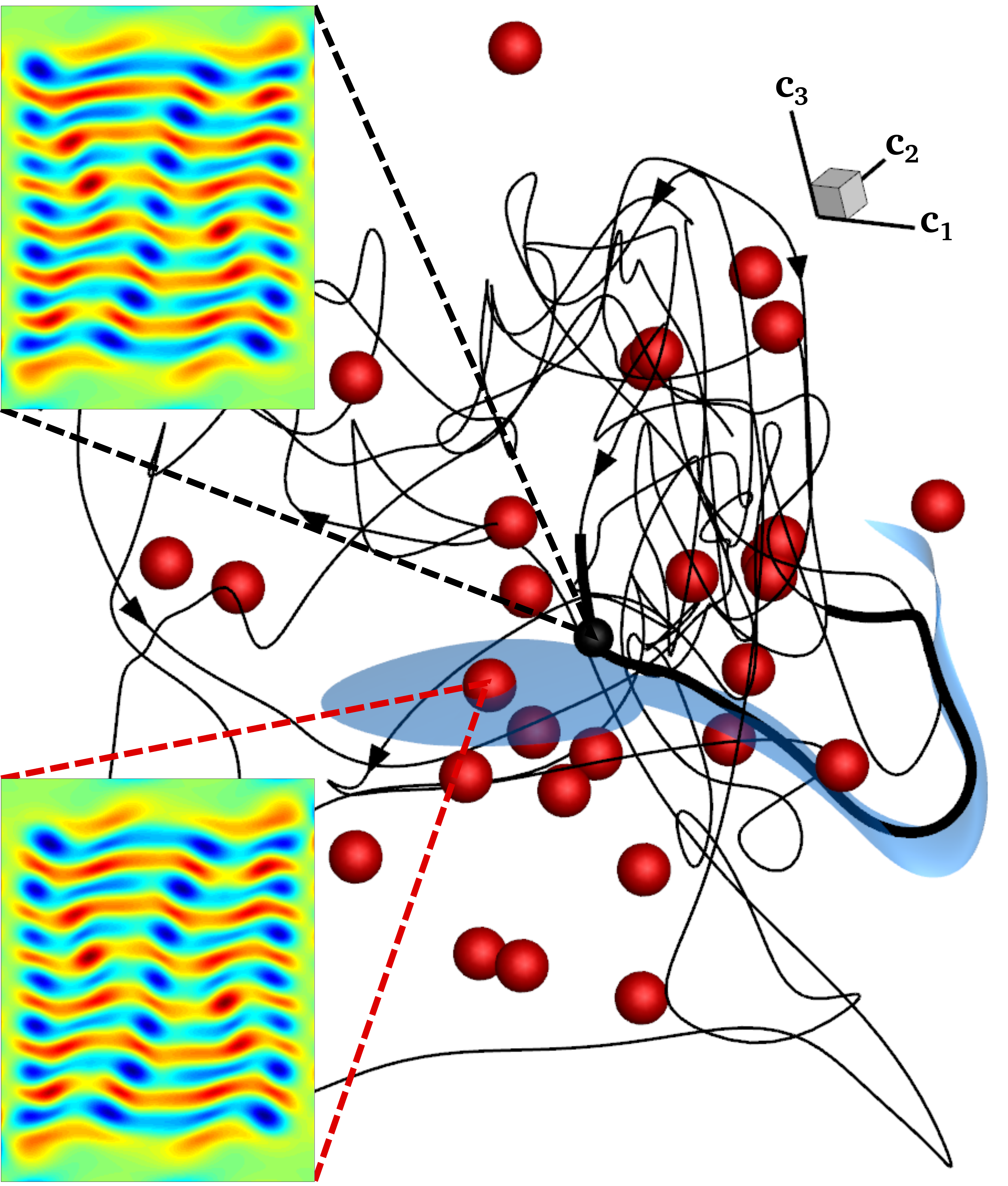}}
\end{center}
\caption{\label{fig:state_space_intro} 
{Low-dimensional projection of the state space (cf. Appendix \ref{sec:projections}).
Black curve is a trajectory describing the temporal evolution of a turbulent flow and red spheres represent different ECSs (equilibria in this particular case).
Thick black line shows the portion of a turbulent trajectory shadowing (a portion of) the unstable manifold (blue surface) of an ECS.
The flow (vorticity) fields in the physical space, corresponding to a particular point on the turbulent trajectory (black ball) and the nearest ECS, are shown on the left.}}
\end{figure}

The dynamical role of ECSs in turbulence is best illustrated using a geometrical description \cite{hopf_1948}, where the flow field in physical space at any given instant is represented as a single point in a high-dimensional state space, as illustrated by \reffig{state_space_intro}. 
The evolution of the turbulent flow corresponds to a {tortuous} trajectory this point traces out in the state space \cite{gibson_2008}.
An infinite hierarchy of ECSs is conjectured to exist in regions of state space explored by turbulent trajectories. 
Near an ECS, the turbulent flow mimics both its spatial and temporal structure \cite{gibson_2008, cvitanovic_2010, suri_2017a}, which is referred to as shadowing. 
However, being unstable, ECSs are visited only fleetingly \cite{kerswell_2007, suri_2017a},
and turbulent flow patterns never become identical to those corresponding  to ECSs.

The geometry of state space around ECSs appears to be shaped by their invariant manifolds \cite{wang_2007, gibson_2008, burak_2017}: turbulent trajectories approach an ECS following its stable manifold and depart following its unstable manifold \cite{suri_2017a}, 
at least when {the spectrum associated with} that ECS is not strongly nonnormal \cite{tobias_2017,farano_2018}.
{Finally, heteroclinic (homoclinic)  connections -- which originate at one ECS 
and terminate at another (the same) ECS -- are conjectured to guide turbulent trajectories between neighborhoods of ECSs \cite{gibson_2008,halcrow_2009,riols_2013}.
We will refer to both heteroclinic and homoclinic connections collectively as dynamical connections.}
In summary, turbulence can be viewed as a deterministic walk between neighborhoods of ECSs, guided by their invariant manifolds and {dynamical} connections.

Numerical simulations of three-dimensional (3D) wall-bounded shear flows, such as plane Couette \cite{nagata_1997}, channel \cite{waleffe_2001}, and pipe flows \cite{faisst_2003, wedin_2004} at moderate Reynolds numbers $Re$ provide strong support for the dynamical relevance of ECSs. 
Equilibrium (EQ) and traveling wave (TW) solutions \cite{nagata_1997, waleffe_2001, faisst_2003, wedin_2004} computed for minimal flow domains \cite{hamilton_1995} capture prominent spatial features (streamwise rolls and streaks) of near-wall coherent structures observed in experiments \cite{kline_1967}.
Certain periodic orbit (PO) and relative periodic orbit (RPO) solutions \cite{kawahara_2001, viswanath_2007} have been shown to describe self-sustained processes \cite{hamilton_1995,waleffe_1997} responsible for destruction and reformation of near-wall coherent structures. 
Statistics of turbulent flows have also been found to agree well with those estimated using only a few ECSs \cite{kawahara_2001,viswanath_2007}, suggesting statistical measures computed using a sufficiently large set of ECSs may indeed converge to those estimated using turbulent time series, providing a direct connection between dynamical and statistical description of turbulence.

While the framework has been developed primarily using numerical simulations, experimental evidence for the existence and dynamical relevance of ECSs in 3D shear flows remains scarce. 
Typically, in pipes and channels, the flow is measured using stereoscopic particle image velocimetry (PIV) within a planar cross section normal to the direction of mean flow. Taylor's hypothesis is then invoked to reconstruct spatially resolved 3D velocity fields as flow structures are advected past the imaging plane. 
Using this technique, flow fields resembling TW solutions {\cite{faisst_2003,wedin_2004, schneider_2007b}} were identified in pipe \cite{hof_2004, lozar_2012, dennis_2014} and channel \cite{lemoult_2014} flow experiments. 
Taylor's hypothesis, however, breaks down where the mean flow is slow (e.g., near stationary walls), and no direct measurements of spatially and temporally resolved 3D velocity fields in experiments have been reported so far.

While numerous ECSs in various 3D shear flows have been computed, very few studies \cite{kerswell_2007,burak_2017} have tested how closely turbulent flows approach ECSs. Those studies, however, were limited to direct numerical simulations in short (less than 10 diameters long) pipes with periodic boundary conditions.
Kerswell \etal \cite{kerswell_2007} have shown that turbulent flow at $Re=2400$ was found in the neighborhoods of TW solutions with $m-$fold ($m = 2,3,4$) symmetry for about 10\% of the time. 
An upper bound of about 20\% for TWs at similar $Re$ was suggested by Schneider \etal \cite{schneider_2007a}.
More recently, Budanur \etal \cite{burak_2017} have tested how closely an RPO, the type of ECS conjectured to be dynamically more relevant than TWs, was shadowed by turbulent trajectories. 
No estimates are currently available for how closely turbulent trajectories in experiments approach ECSs.

The majority of the studies exploring invariant manifolds \cite{gibson_2008} focused on their role in direct laminar-turbulent transition \cite{kerswell_2005,eckhardt_2009} in 3D shear flows. 
Several TWs \cite{itano_2001,wang_2007,kerswell_2007,schneider_2007b,viswanath_2009,burak_2018} 
and POs \cite{kawahara_2001,toh_2003,Kawahara_2005,viswanath_2007,halcrow_2009,duguet_2008a} were found to lie on the laminar-turbulent boundary. 
The boundary itself can be constructed as the union of stable manifolds of these ``edge states'' and determines which nearby state space trajectories become turbulent and which relaminarize. 
However, the dynamical relevance of invariant manifolds in sustained 3D turbulence -- either in simulation or experiment -- is yet to be verified.

The dynamical relevance of ECSs (unstable equilibria) and their invariant manifolds, however, was recently established by Suri \etal \cite{suri_2017a} in the context of  quasi-two-dimensional (Q2D) turbulence. 
Q2D flows, generated in shallow electrolyte layers driven by a horizontal electromagnetic force, are often studied as models of geophysical flows \cite{vanheijst_2009, dolzhansky_2012}. 
In experiments, spatially and temporally resolved velocity fields can be easily measured (at the electrolyte-air interface) using two-dimensional (2D) PIV, while the flow can be modeled using a strictly 2D equation \cite{suri_2014}.
This allowed Suri \etal to compute 16 unstable equilibria of a Q2D Kolmogorov-like flow, directly using PIV data. 
Furthermore, it was shown that the evolution of turbulent trajectories, in both simulation and experiment, in the neighborhood of an ECS can be forecast by constructing its unstable manifold. 

The goal of this article is to address some open questions regarding the dynamical role of unstable equilibria and their invariant manifolds, once again using the Q2D Kolmogorov-like flow as the test bed.
Specifically, we show that turbulent flow in both the experiment and simulations comes quite close to almost all the unstable equilibria of the model that have been computed so far.
Using a pair of representative equilibria we illustrate that, after turbulent trajectories enter the neighborhood of these ECSs they closely shadow the corresponding unstable manifolds for an extended period of time. 
The article is structured as follows: In sections  \ref{sec:exp} and \ref{sec:sim} we present a brief overview of the experimental setup and numerical simulations, respectively. 
Our results are presented in Section \ref{sec:results} and conclusions in Section \ref{sec:summary}.

\section{Experimental Setup}\label{sec:exp}

The Kolmogorov flow used in classical investigations of hydrodynamic stability \cite{meshalkin_1961,chandler_2013} refers to a strictly 2D flow driven by a sinusoidal forcing. Such a flow, however, is an idealization that is impossible to reproduce exactly in experiment.
We will describe here an experimental setup that preserves, to the extent possible, its key features. 

\begin{figure}
\begin{center}
\subfloat[]{\includegraphics[width=3in]{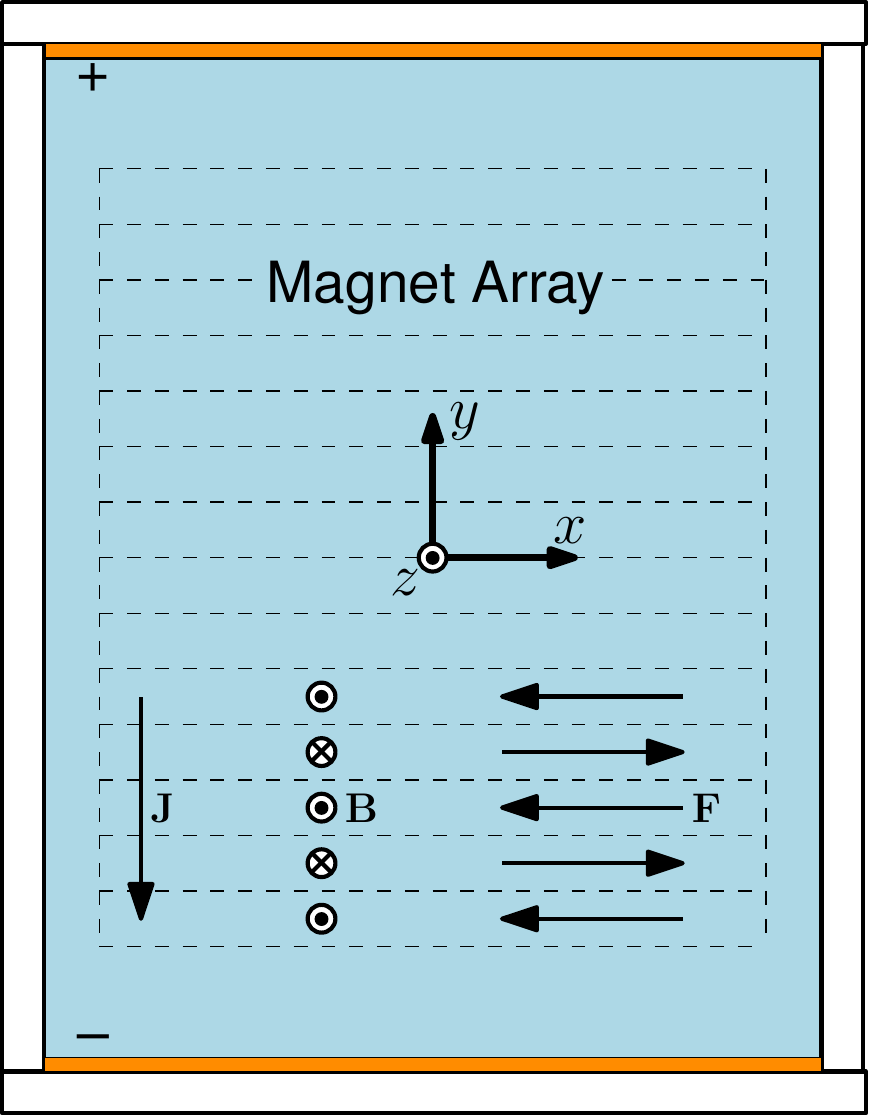}} \\
\subfloat[]{\includegraphics[width=3in]{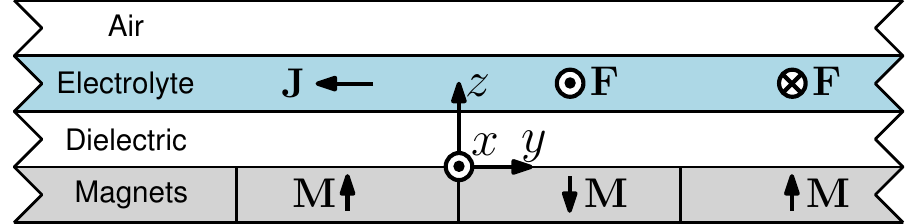}}
\end{center}
\caption{\label{fig:exp_setup}
Schematic of the experimental apparatus. (a) Top view showing a magnet array with adjacent magnets (dashed lines) having magnetization in opposite directions (up/down/up/...). 
The array is placed over an aluminum plate and a container is constructed using acrylic end walls and copper electrode side walls. (b) Side view showing the cross section of the apparatus holding two immiscible fluid layers, a heavy dielectric and a lighter electrolyte. 
A direct current density ${\bf J}$ passing through the electrolyte interacts with the magnetic field ${\bf B}$ produced by the magnet array, exerting the Lorentz force ${\bf F}={\bf J}\times{\bf B}$. 
Spatiotemporally resolved 2D velocity fields are obtained using particle image velocimetry by imaging the tracer particles at the electrolyte-air interface with a camera (not shown) suspended above the container.} 
\end{figure}

To create a nearly sinusoidal forcing, we arrange 14 NdFeB magnets (grade N42) to form an array of dimensions 15.24\si{cm}$\times$($14\times1.27$\si{cm})$\times$0.32\si{cm}, as shown in \reffig{exp_setup}(a). Adjacent magnets {have opposite polarity,} 
resulting in a nearly sinusoidal magnetic field {${\bf B} \approx B_z\hat{\bf z}$ with $B_z \approx e^{-\kappa z} \sin(\kappa y)$} at the center of the array. Here $\kappa = \pi/w$ and $w=1.27$\si{cm} is the width of each magnet. 
The magnet array is placed on a horizontal aluminum plate and is padded with 0.32\si{cm} thick aluminum bars to create a flat surface.   
A  50 $\mu$m-thick black contact paper is placed over this flat surface to provide uniform black background for better flow visualization.
Acrylic bars and electrodes, running parallel to $y$ and $x$ axes respectively, are glued on top of contact paper to create a rectangular container of dimensions 17.8\si{cm}$\times$22.9\si{cm} with the magnet array at its center.

To create a nearly 2D flow, we use the two-immiscible-fluid layer configuration shown in \reffig{exp_setup}(b). The bottom layer is a dielectric (perfluorooctane with density $\rho=1769$ kg/m$^3$ and viscosity $\mu= 1.3$ mPa$\cdot$s) and the top one is an electrolyte (1 M CuSO$_4$ with 40\% glycerol by weight with density $\rho=1160$ kg/m$^3$ and viscosity $\mu=5.7$ mPa$\cdot$s); each layer is 0.3\si{cm} thick. 
Passing a uniform direct current density ${\bf J} = J\hat{\bf y}$ through the electrolyte generates a nearly sinusoidal Lorentz force ${\bf F} \approx {J} e^{-\kappa z}\sin(\kappa y)\uv{x}$ that drives the flow in both layers. 
The flow in the experiment is visualized by seeding the electrolyte-air interface with Glass Bubbles (K15) manufactured by 3M.  
Spatiotemporally resolved images of the entire lateral extent of the flow are recorded at 15 Hz using the DMK 31BU03 camera which has a 1024$\times$768 pixel CCD sensor.  
Velocity field ${\bf u}(x,y,t)$ at the electrolyte-air interface is calculated using the Prana PIV package \cite{prana} {employing} the ``Deform" multigrid PIV algorithm. The grid resolution of the resulting PIV measurements is about 120 $\times$ 160, or approximately 9 grid points per magnet width. 

The dynamical regimes in the experiment are parametrized using the Reynolds number
\begin{equation}\label{eq:def_re}
Re = \frac{\bar{u}w}{\bar \nu}.
\end{equation} 
Here, $\bar{\nu} = 3.26$ mm$^2$/s is the depth-averaged kinematic viscosity of the two fluid layers \cite{suri_2014} and the characteristic velocity $\bar{u}$ is defined as the spatial root-mean-square average of ${\bf u}$ over the central $8w\times8w$ region which is subsequently temporally averaged over the entire time-series.

\section{Theoretical Model}\label{sec:sim}

The evolution of the flow is modeled using a strictly 2D equation \cite{suri_2014}
\begin{equation}\label{eq:2dns_mod_ndim}
\frac{\partial{\bf u}}{\partial t} + \beta {\bf u}\cdot\nabla{\bf u} = -\nabla p + \frac{1}{Re}\left(\nabla^2 {\bf u}- \gamma{\bf u}\right) + \tribra{{\bf F}_{\|}}_z,
\end{equation} 
derived from first principles by averaging the 3D Navier-Stokes equation along the confined ($z$) direction.
In the above equation $\tribra{{\bf F}_{\|}}_z$ is the depth-averaged 2D force density while  
$p$ is the 2D kinematic pressure. 
The velocity field ${\bf u} = (u_x,u_y)$ in the 2D model is assumed to be incompressible ($\nabla\cdot{\bf u}=0$) which is a good approximation for $Re \leq 40$ \cite{suri_2014}. 
In the experiment, the solid boundary at the bottom causes a {vertical} gradient in the magnitude of the horizontal velocity
\cite{satijn_2001, suri_2014}.
Equation \eqref{eq:2dns_mod_ndim} {describes the} change in inertia of the fluid layers due to this gradient using prefactor $\beta<1$ to the nonlinear term {and the term $-\gamma {\bf u}$ represents the associated viscous shear stresses in the bottom fluid layer.} 
For the fluid layer configuration in our experiment, $\beta = 0.83$ and $\gamma = 3.22$ deviate significantly from values corresponding to a strictly 2D flow ($\beta=1$, $\gamma = 0$). 
Lastly, equation \eqref{eq:2dns_mod_ndim} was non-dimensionalized by choosing $w$ (magnet width), $\bar{u}$, $w/\bar{u}$, and $\bar{u}^2$ as scale factors for length, velocity, time, and kinematic pressure $p$, respectively \cite{tithof_2017}.

While the forcing profile near the center of the magnet array in experiment is sinusoidal, it is fairly complicated near the lateral boundaries. 
To accurately replicate such forcing profile, {the effective forcing} $\tribra{{\bf F}_{\|}}_z$  in the 2D model is computed following a first principles approach. 
The magnet array in the experiment is modeled as a 3D lattice of uniformly magnetized dipoles, with dipoles in adjacent magnets pointing oppositely along $\uv{z}$ and $-\uv{z}$. 
The net magnetic field $B_z(x,y,z)$  {produced by the dipole lattice is calculated and the 3D Lorentz force density ${\bf F}=JB_z\uv{x}$ is depth-averaged to obtain  the 2D force density} \cite{tithof_2017}.

{In the case of uniform magnetization, the  forcing profile is anti-symmetric under the coordinate transformation $\rsym:(x,y)\rightarrow(-x,-y)$, i.e., $\rsym\tribra{{\bf F}_{\|}}_z = -\tribra{{\bf F}_{\|}}_z$. $\rsym$ is equivalent to rotation by $\pi$ about the $z-$axis passing through the center of the domain. 
Under  $\rsym$, the velocity field transforms as $\rsym {\bf u}(x,y) \rightarrow -{\bf u}(-x,-y)$ which leads to \refeqs{2dns_mod_ndim} being equivariant under ${\rsym}$.}
However, since the magnets used in the experiment are not uniformly magnetized {and the bottom of the container is not perfectly horizontal},  $\rsym$ is weakly broken.

Direct numerical simulations based on the 2D \refeqs{2dns_mod_ndim} in its semi-discrete form \cite{leveque_1992} are performed on a computational domain with lateral dimensions and no-slip velocity boundary conditions identical to those in the experiment. 
Velocity and pressure fields are spatially discretized on a 2D marker and cell (MAC) staggered grid of dimensions 280$\times$360. 
This corresponds to a resolution of 20 cells per magnet width  with grid spacing $\delta x$=$\delta y$=0.05. 
Spatial derivatives in \refeqs{2dns_mod_ndim} are approximated using finite differences: the 2D Laplacian operator with a five-point central difference formula and the nonlinear term with a modified MAC formula \cite{griebel_1998}.
Temporal integration of \refeqs{2dns_mod_ndim} is performed using the P2 projection scheme to enforce incompressibility of the velocity field at each time step \cite{vankan_1986, armfield_1999}. 
Temporal update of linear terms uses the implicit Crank-Nicolson scheme and that of the nonlinear term uses the explicit Adams-Bashforth scheme \cite{armfield_1999}. 
For all numerical data presented in this article, a time step $\delta t = 1/110$ was used for temporal integration to ensure the CFL number ${\max\{{u_x}, u_y\}\delta t/\delta x} \leq 0.5$. 
More details regarding numerical simulations can be found in Suri \etal \cite{suri_2017a} and Tithof \etal \cite{tithof_2017}.

Since the lateral dimensions and boundary conditions in the experiment and the simulation are identical  {flow fields} from the experiment, {satisfying certain specific criteria described below, can be used to compute dynamically relevant ECSs}. 
To facilitate this, the PIV  measurements on the coarser $120\times 160$ grid are interpolated onto the finer $280\times 360$ simulation grid. 
The interpolated fields (${\bf u}_{*}^{exp}$) are then projected onto the divergence-free subspace employing Helmholtz-Hodge decomposition  to obtain initial conditions ${\bf u}_{ic}^{exp}$ for the simulations \cite{suri_2017b}. 
The maximum difference between ${\bf u}_{*}^{exp}$ and ${\bf u}_{ic}^{exp}$, computed as $\|{\bf u}_{ic}^{exp}-{\bf u}_{*}^{exp}\|/\|{\bf u}_{*}^{exp}\|$, across all initial conditions tested, was less than 0.025. This confirms the flow at the electrolyte-air interface is nearly incompressible at $Re=22.5$ (the Reynolds number considered herein). 

\section{Results}\label{sec:results}

A quantitative study of the transition from laminar flow to turbulence in this system (using both experiment and numerical simulations) has recently been performed by Tithof \etal\cite{tithof_2017}. 
As $Re$ increases, the flow undergoes a sequence of bifurcations and becomes weakly turbulent at $Re\approx 18$. 
In the remainder of this paper, we will focus on the dynamics and the role of ECSs at $Re=22.5$. 
To illustrate the characteristic dynamics and flow structures at this $Re$, we have included videos showing evolution of vorticity fields from experiment and simulation in supplemental material (videos 1 and 2).
Temporal auto-correlation computation shows that the correlation time at this $Re$ is $\tau_c = 27\pm1$\si{s}  (cf. Appendix \ref{sec:sec_auto_corr}). 
In the experiment, five separate  $125\tau_c$-long (3600\si{s})  runs were performed to search for signatures of invariant solutions.
In addition, two $1000 \tau_c$-long (28000\si{s}) time-series were generated using numerical simulations. 
Velocity fields in both simulations and experiment were sampled at uniform intervals of $\Delta t = 0.037\tau_c$ (1\si{s}) for analyzing the dynamics. 

A leading theory assumes that dynamically dominant ECSs underlying fluid turbulence correspond to temporally periodic solutions of the governing equations \cite{cvitanovic_1988,kawahara_2001, cvitanovic_2013,budanur_2017} and there is some evidence supporting this assumption.
In particular, a previous numerical investigation \cite{chandler_2013} of a 2D  Kolmogorov flow (described by equation \eqref{eq:2dns_mod_ndim} with $\beta=1$ and $\gamma=0$) with periodic lateral boundary conditions found many tens of (relative) temporally periodic solutions visited by turbulent dynamics. 
Unlike the flow on a periodic domain, which has both continuous and discrete symmetries \cite{chandler_2013, tithof_2017}, the laterally bounded system here has only a discrete symmetry $\rsym$ and so it has no relative solutions.

\begin{figure}[!t]
\begin{center}
\includegraphics[width=3.3in]{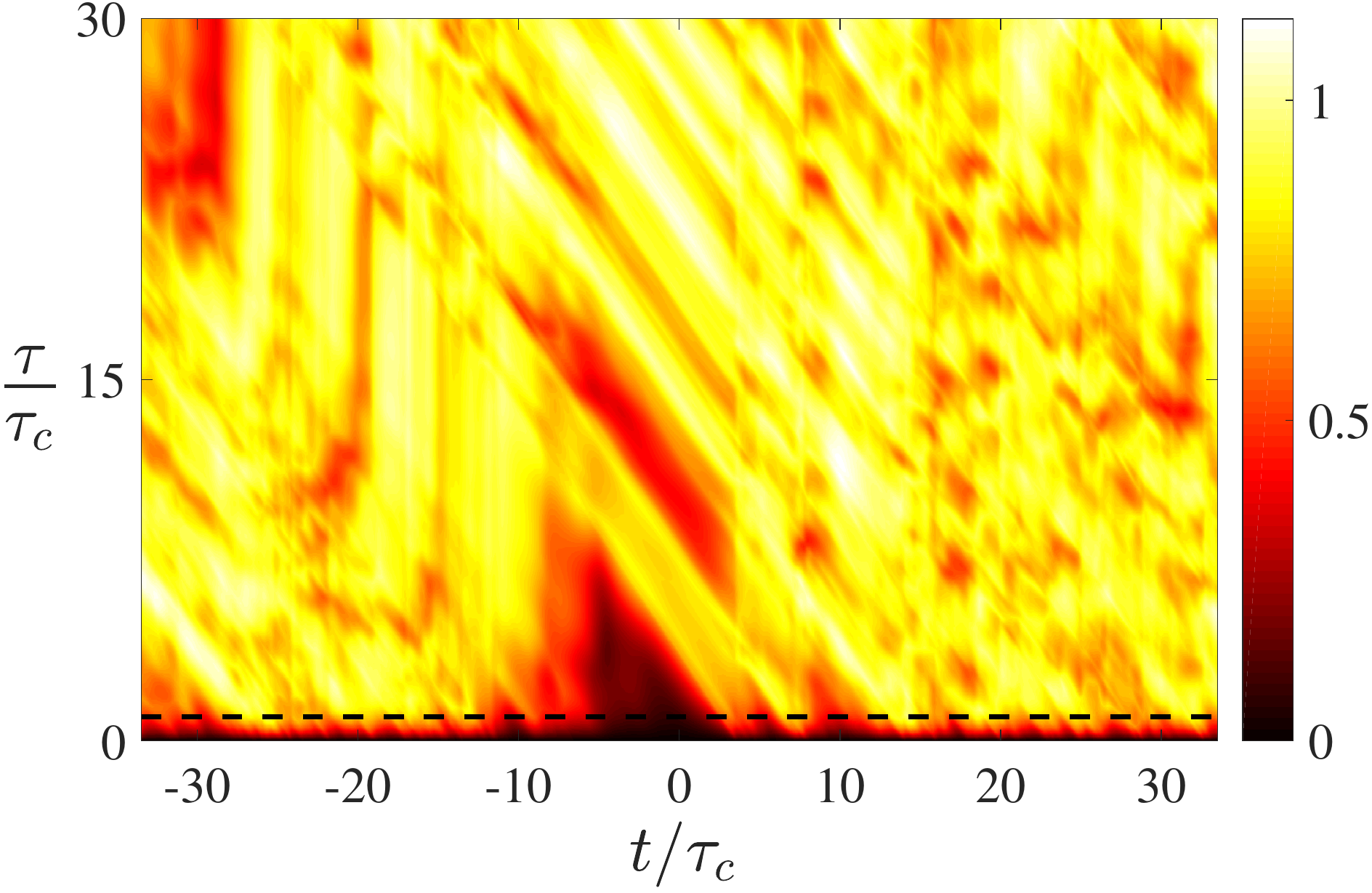}
\end{center}
\vspace{-4mm}
\caption{\label{fig:rec_plots} {A sample recurrence plot used to identify signatures of ECSs in weakly turbulent flow produced by our simulations.} {Low (high) values of $R(t,\tau)$ indicate that flow fields at two instants $t$ and $t+\tau$ are similar (different).}
The dashed line corresponds to the correlation time $\tau_c$.}
\end{figure}

To identify signatures of time-periodic solutions guiding turbulent flow we perform recurrence analysis \cite{eckmann_1987} of the velocity field time series. 
A recurrence diagram, like the one shown in \reffig{rec_plots}, represents a measure of how similar the flow field at an instant $t$ is compared with the flow field at a later instant $t+\tau$, quantified using the function
\begin{equation}\label{eq:eq_rec}
R(t,\tau) = \min_g\frac{\|{\bf u}(t)-g{\bf u}(t+\tau)\|}{\|{\bf u}(t)\|},
\end{equation}
where $\tau>0$ and $g=\{1,\rsym\}$ are the elements of the group of discrete symmetries of the flow. While turbulent flow fields  are not invariant under $\rsym$, a flow field {${\bf u}$ and its rotated version $\rsym{\bf u}$} can recur in time with equal (comparable) likelihood in the simulation (experiment). 
Equation \eqref{eq:eq_rec} accounts for this by identifying the closest recurrence among symmetry related copies.

In \reffig{rec_plots}, a region representing a low value of $R(t,\tau)$ indicates a near-recurrence, which means that flow fields at instants $t$ and $t+\tau$ are similar. 
Deep local minima at $\tau>\tau_c$ {(dark red islands)} suggest that turbulent flow shadows a periodic orbit with temporal period $T\approx\tau$.  
To our surprise, both in simulation and experiment, very few ($\leq10$) such near-recurrences were observed, suggesting time-periodic solutions are not the most dynamically important ECSs at $Re=22.5$. 

Instead, the most prominent feature was the presence of {dark red} triangular regions with their base at $\tau = 0$ and height comparable to $\tau_c$, like the ones at $t\approx-10,0,5,10$ in \reffig{rec_plots}. 
Such features reflect significant slowing down of the evolution during which the flow becomes nearly time-independent.
This is a signature of the turbulent trajectory visiting the neighborhood of an unstable equilibrium solution.
This inference can be rationalized using the analogy of a rotating pendulum slowing down near its inverted position, which corresponds to an unstable equilibrium. 
The discussion hereafter will focus on identifying such unstable equilibria, their stability properties, and the dynamical role they play in shaping the evolution of nearby turbulent trajectories.

\subsection{Unstable Equilibrium Solutions} \label{sec:unstable_equilibria}

Unlike the inverted pendulum example, unstable equilibria of the governing equations describing the flow considered here are not known \textit{a priori}. 
Hence, we hypothesize that {the turbulent trajectory passes close an equilibrium when it is evolving sufficiently slowly.}
To identify such instants, we define the instantaneous {state space} speed $s(t)$ {\cite{suri_2017a, acharya_2017}} of the turbulent trajectory as the {properly normalized} rate of change in velocity fields
\begin{equation}\label{eq:sss_sim}
s(t)=\frac{\tau_c}{\|{\bf u}(t)\|}\left\|\frac{d{\bf u}}{dt} \right\| 
\approx \frac{\tau_c}{\Delta t}\frac{\left\|{\bf u}(t+\Delta t) - {\bf u}(t)\right\|}{\|{\bf u}(t)\|}.
\end{equation} 
In the above equation, $\Delta t = 0.037\tau_c$ (1\si{s}) is the interval between successive samples of velocity fields in both simulation and experiment. 

\begin{figure}[!t]
\centering
\includegraphics[width=3.3in]{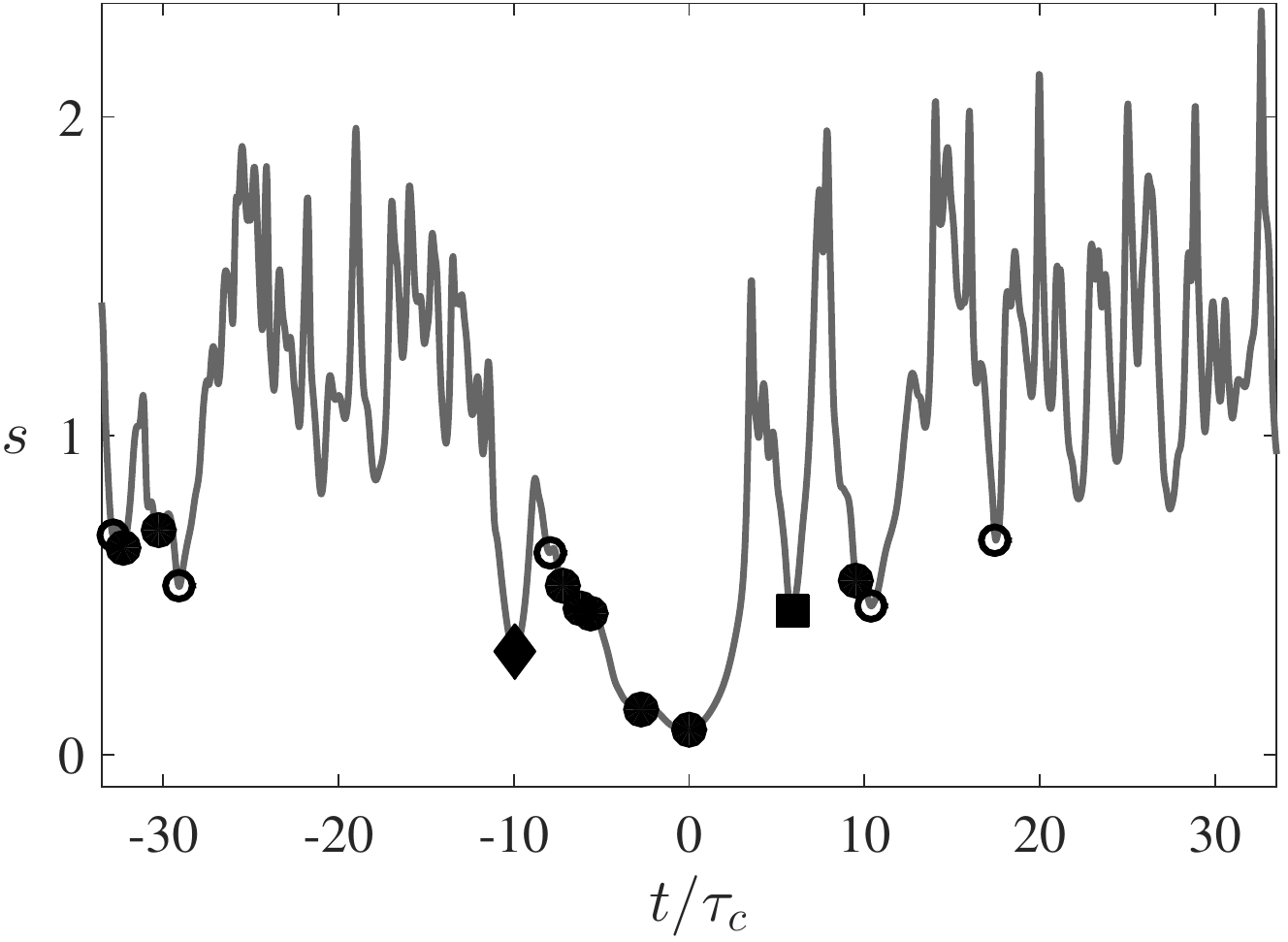}
\caption{\label{fig:sss_sim} State space speed $s(t)$, with symbols indicating the deep local minima. Filled symbols designate a nearby unstable equilibrium was computed when the Newton-Krylov solver was initialized using the corresponding turbulent flow field. 
Different filled symbols represent convergence to distinct equilibria (cf. \reffig{sample_sols_sim}). Open circles indicate the solver failed to converge to an equilibrium.} 
\end{figure}

Figure \ref{fig:sss_sim} shows the state space speed for the same segment of turbulent trajectory analyzed in \reffig{rec_plots}.
The position of the deep minima ($s\leq0.7$), identified with symbols, closely corresponds to the location of the prominent {red} triangular regions in the recurrence plot.
{Note that the deepest minimum $\min_t s(t) = 0.08$ is significantly lower than the temporal average of $s$, which is nearly unity.}
To determine whether these minima are associated with a nearby unstable equilibrium, we initialized a Newton-Krylov solver \cite{kelley_2003,mitchell_2013} using the turbulent flow fields which correspond to the respective minima.
In several cases, labeled with filled symbols in \reffig{sss_sim}, the solver identified a nearby unstable equilibrium ${\bf u}_0$. 
For some initial conditions ${\bf u}_{ic}$, however, the solver failed to converge to an equilibrium solution. 
Several different unstable equilibria were found this way, which we indicated using distinct symbols in \reffig{sss_sim}. 
Here and below, $t=0$ denotes the global minimum of $s(t)$ within the temporal window shown.

\begin{figure}[!t]
\centering
\subfloat{\includegraphics[width=3.25in]{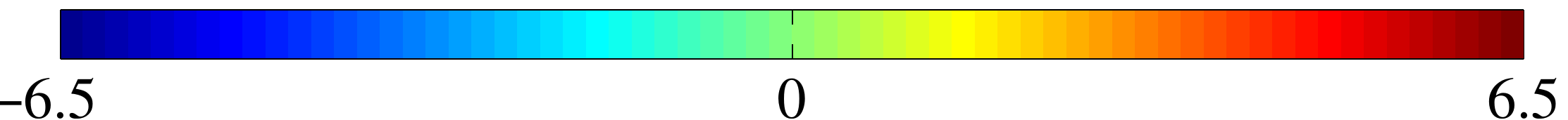}\hspace{1mm}}\\
\vspace{-3mm}
\addtocounter{subfigure}{-1}
\subfloat[]{\includegraphics[height=2in]{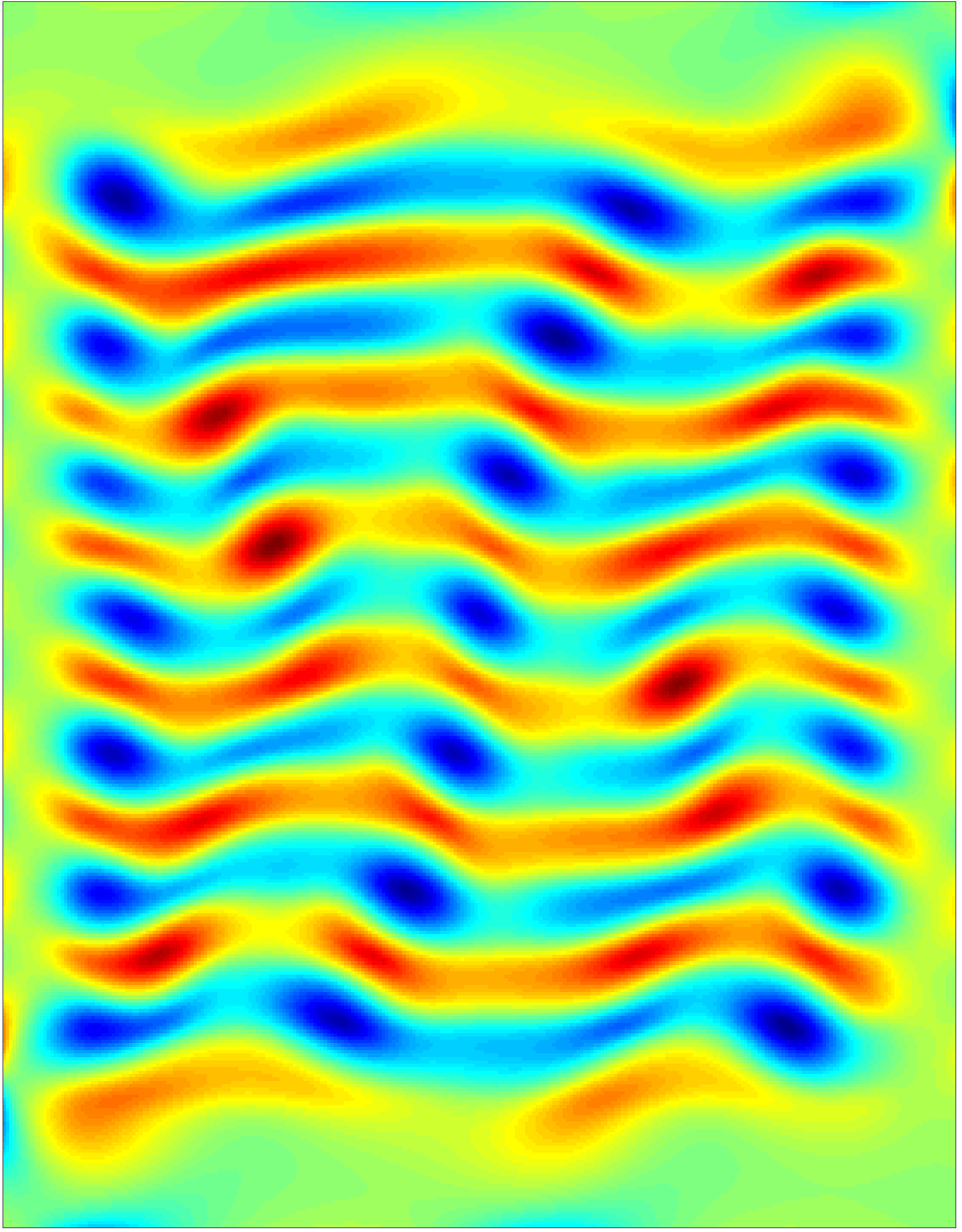} \hspace{1.75mm}
\includegraphics[height=2in]{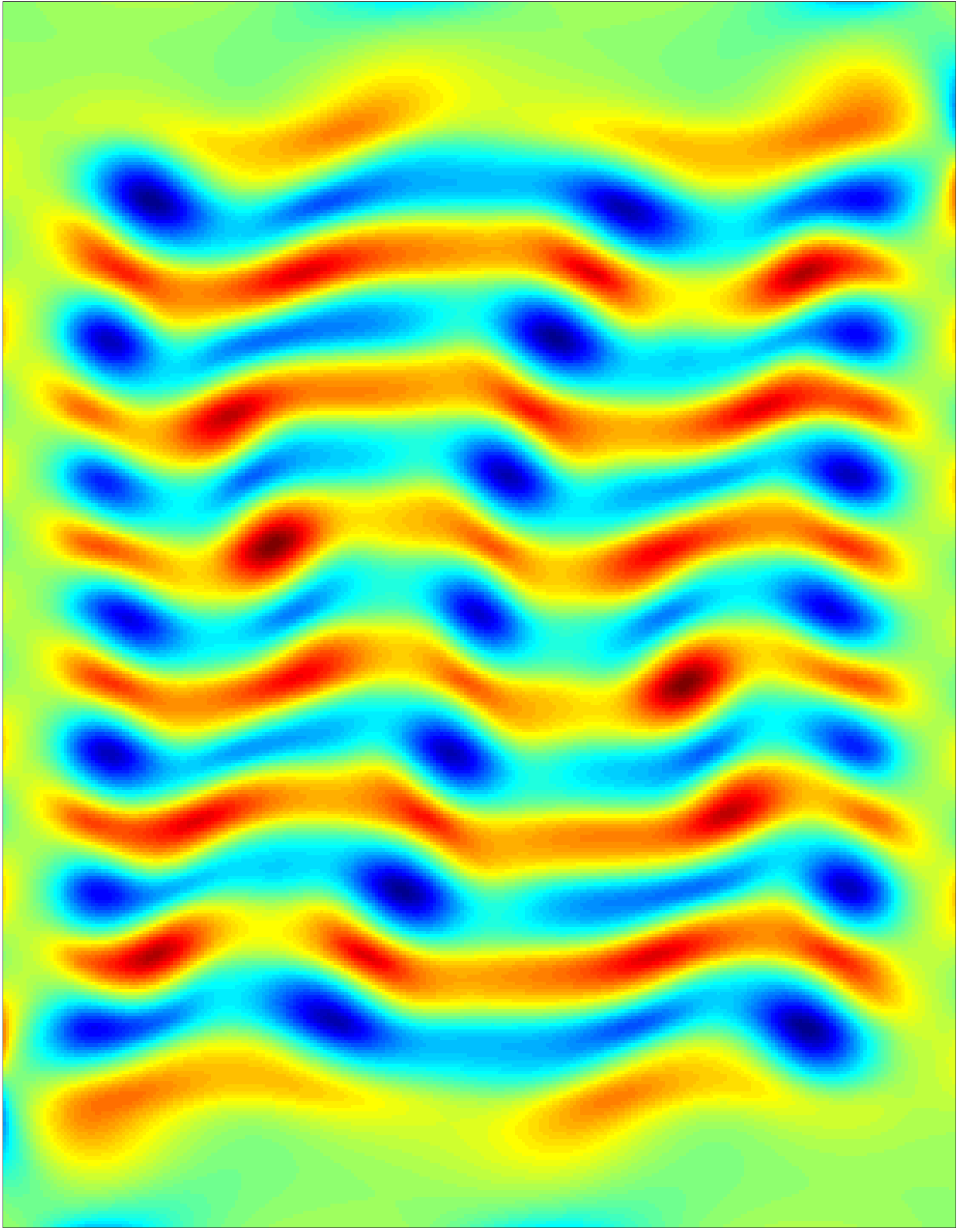}} \\
\vspace{-3mm} 
\subfloat[]{\includegraphics[height=2in]{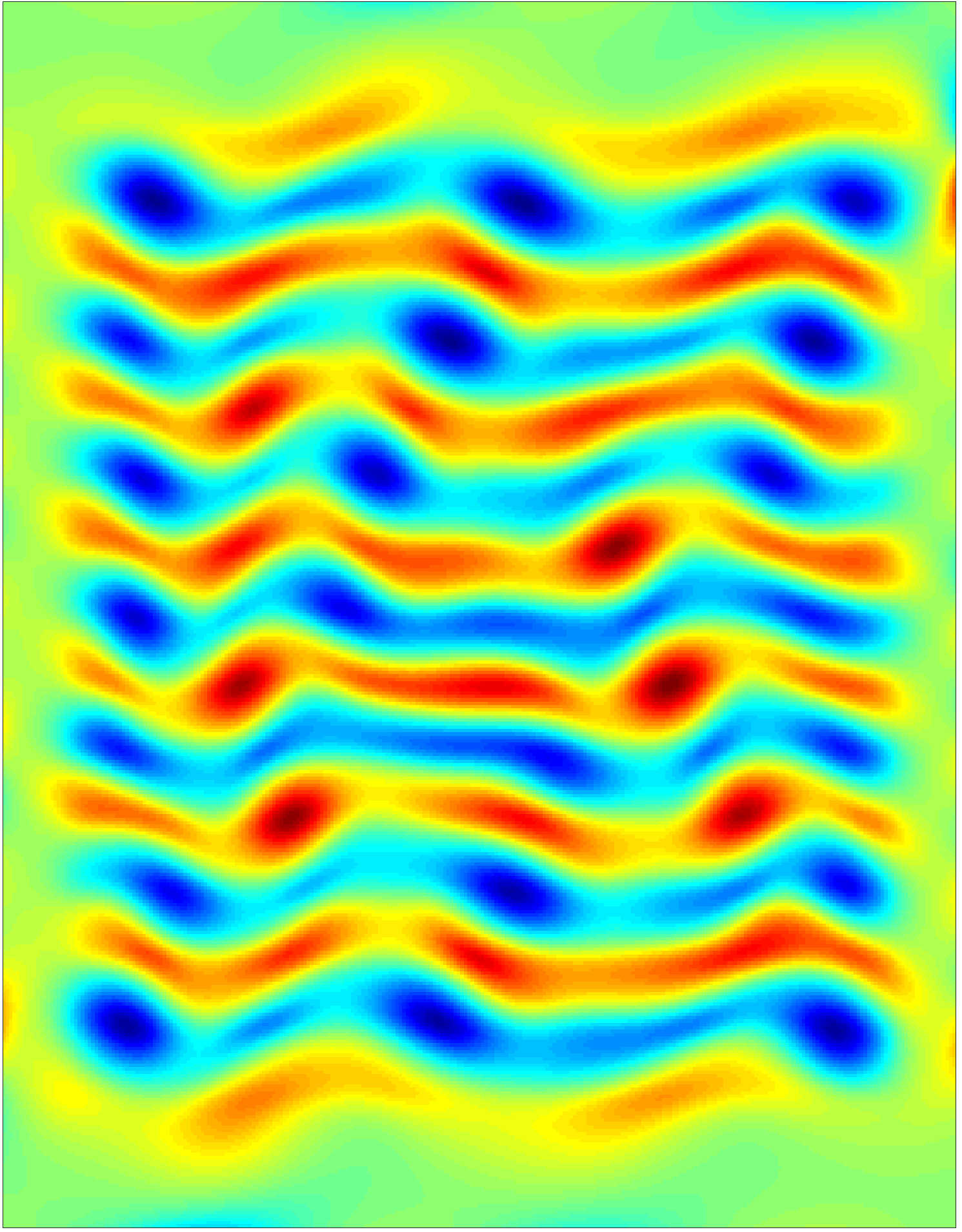} 
\hspace{1.75mm}
\includegraphics[height=2in]{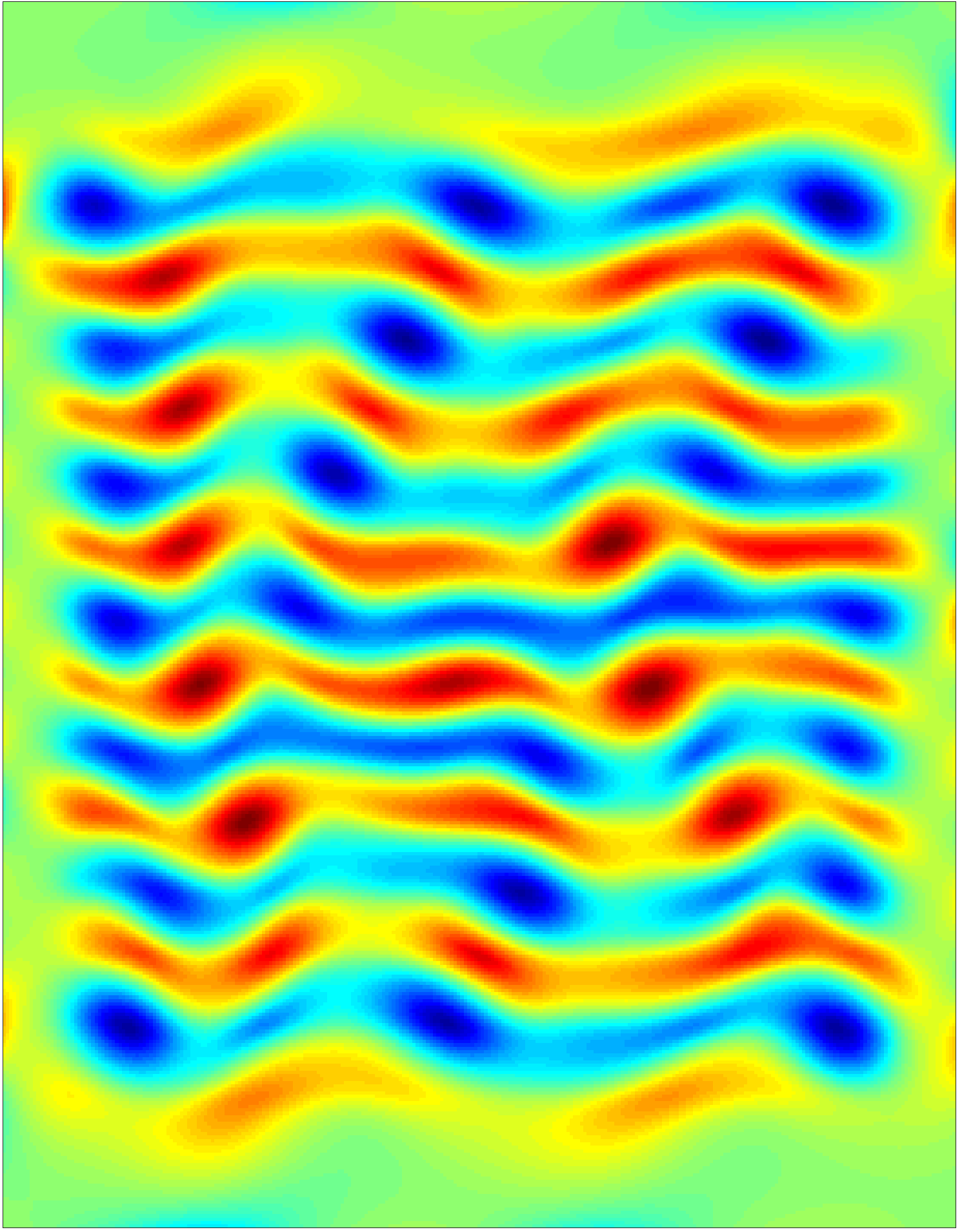}}
\caption{\label{fig:sample_sols_sim} Equilibria (a) E01 and (b) E10 (see Table \ref{table:sol_summary}). 
The left column shows initial conditions from the numerical simulation corresponding to local minima (a) $t=0$ (black circle) and (b) $t =-10$ (black diamond) in \reffig{sss_sim}. 
The right column shows the {corresponding} equilibria. 
The normalized distances from the initial condition{s} 
to E01 and E10 are $D_0^{ic} = 0.20$ and $0.61$, respectively.} 
\end{figure}

The initial conditions, represented by contour plots of vorticity $\omega = (\nabla \times {\bf u})\cdot\uv{z}$, are compared with the respective unstable equilibria of the 2D model in \reffig{sample_sols_sim}.  
The visual similarity of the corresponding flow states in the physical space is striking and unequivocally illustrates that turbulent trajectory passes very close to unstable equilibria, which is consistent with the dramatic slowdown in evolution. 
Notice that, since $\rsym\omega(x,y) \rightarrow \omega(-x,-y)$, {the equilibrium} E01 is invariant under $\rsym$, while E10 is not.
In all, flow fields at 350 deep minima of $s(t)$ were tested for convergence in the simulation and 55 (about 15\%) of these converged to 18 distinct unstable equilibria.
While the temporal window in \reffig{sss_sim} shows a higher percentage of convergence, it was chosen since it includes the deepest minima across all the data.

Failure of the Newton-Krylov solver to converge to an unstable equilibrium from a given initial condition does not necessarily indicate that there is no unstable equilibrium nearby, since convergence is only guaranteed for sufficiently close initial conditions.
It is quite likely that a more robust solver, such as an adjoint-based one \cite{farazmand_2016}, might identify additional unstable equilibria.
It is also worth noting that the success or failure of the Newton-Krylov method is correlated with the value of $s(t)$ at the local minimum (i.e., the success rate is the highest for the deepest minima), but the correlation is  not perfect, as illustrated by \reffig{sss_sim}.
This is not surprising, given the highly anisotropic structure of chaotic sets, such as the one underlying turbulent dynamics of the Kolmogorov flow, in the state space.

\begin{figure}[!t]
\centering
\subfloat[]{\includegraphics[height=2in,angle = 180]{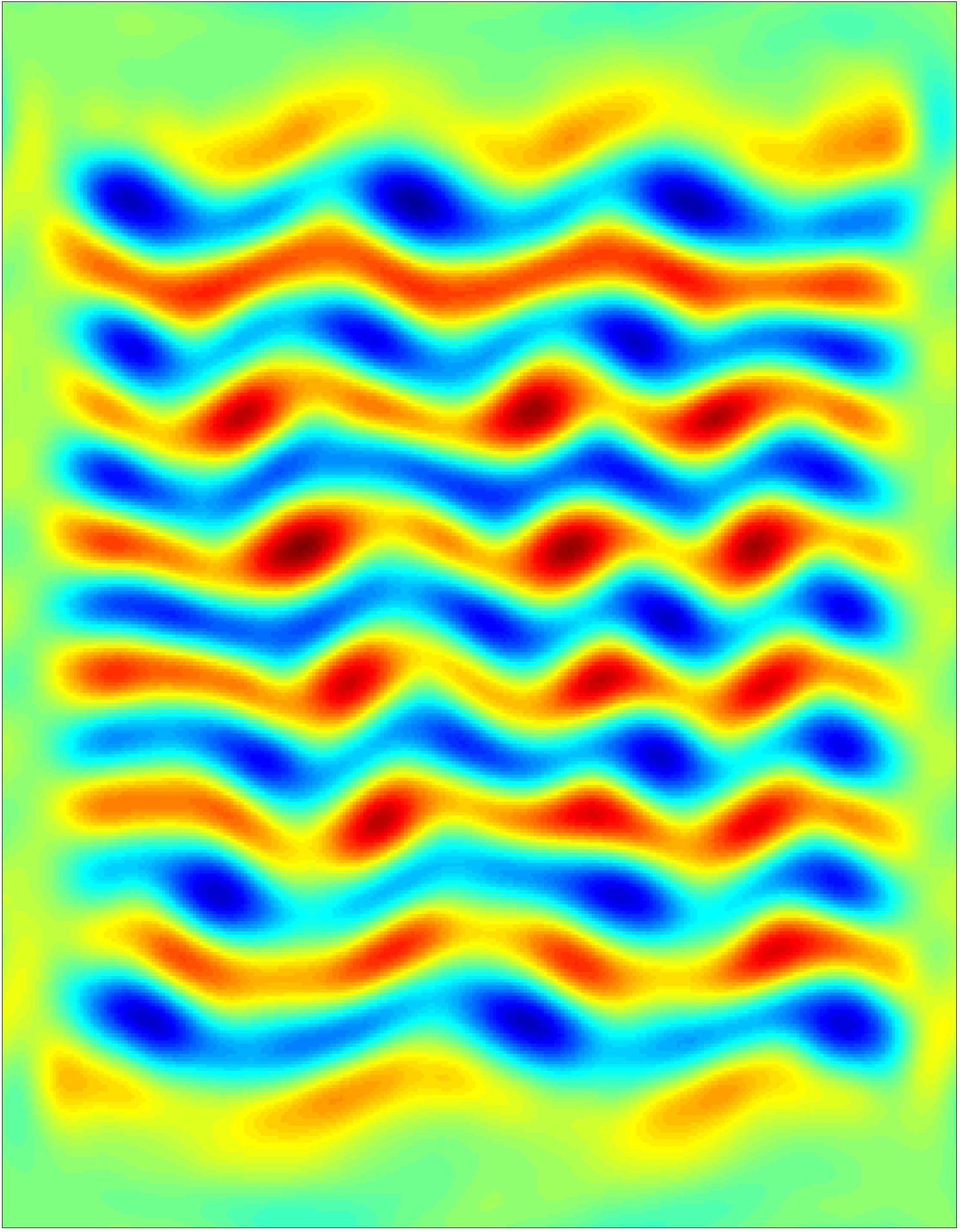}} \hspace{1.75mm}
\subfloat[]{\includegraphics[height=2in,angle = 180]{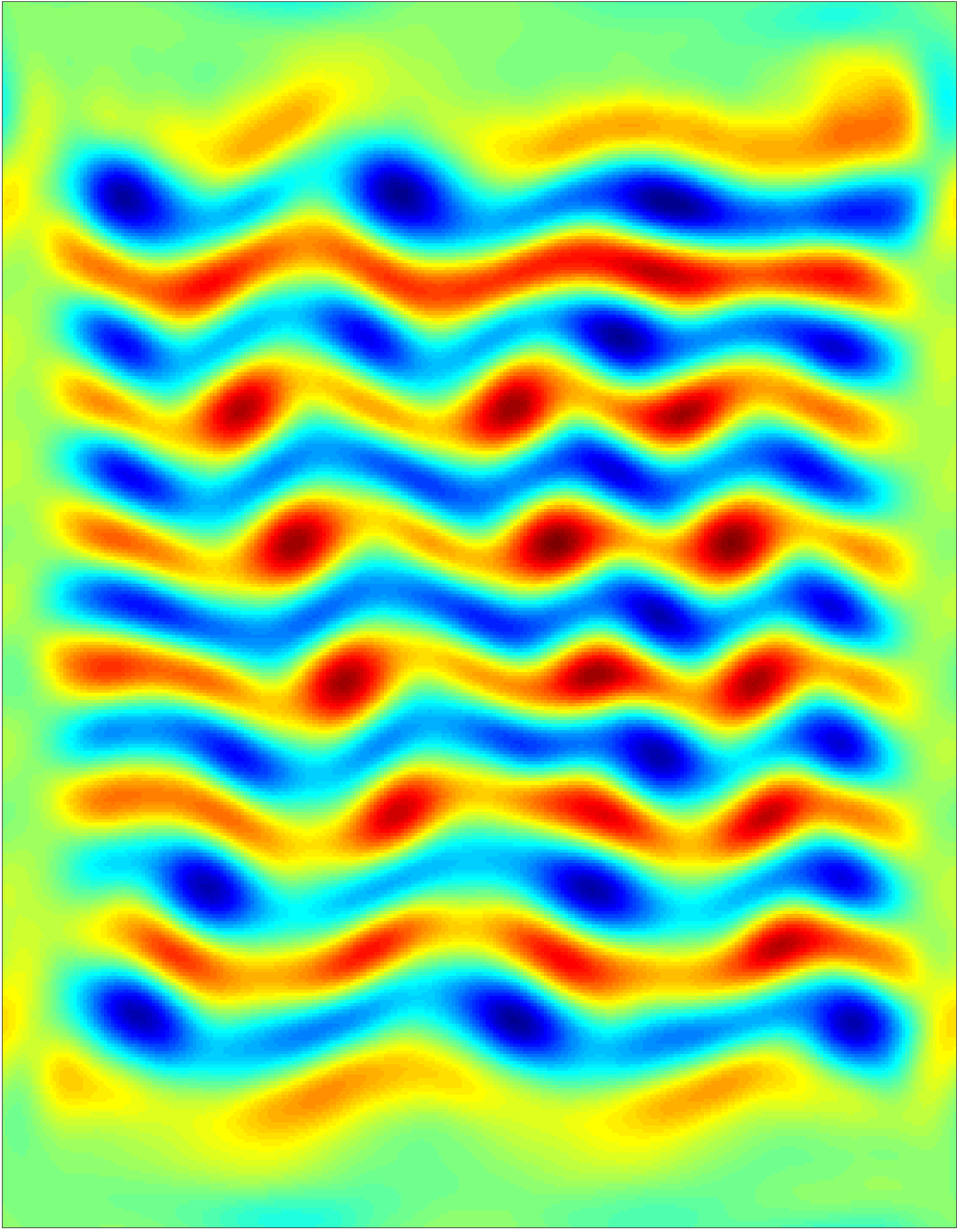}}
\caption{\label{fig:sample_sols_exp} {Equilibrium E20: (a) initial condition from the experiment and (b) the corresponding solution.} 
The normalized difference between the two flow fields is $D_0^{ic} = 0.33$.
The colormap used here and below is the same as in \reffig{sample_sols_sim}.} 
\end{figure}

Following a similar methodology, several equilibria of the 2D model were also computed by initializing the Newton-Krylov solver with processed PIV data from experiment. 
As an illustration, \reffig{sample_sols_exp} shows {the flow field corresponding to a minimum of $s(t)$} and the corresponding equilibrium.
Due to the relatively short duration (125$\tau_c$) of each {experimental run,} a higher threshold $s\leq 1.1$ was chosen to identify a number of minima comparable to that in the simulations.
From about 300 initial conditions from the five experimental runs, 24 (about 8\%) converged to 19 distinct equilibria. 

\setlength{\tabcolsep}{6pt}
\begin{table*}[!htbp]\centering
\ra{1.3}
\begin{tabular}{@{}cc|cc|cc||cccc@{}}
\toprule
{Sol} & $\|{\bf u}_0\|$ &  IC & $\min\limits_{ic}D_{0}^{ic}$ & $\min\limits_{t}D^{sim}_0(t)$ & $\min\limits_{t}D^{exp}_0(t)$ & $N_{{u}}$ & $N_{KY}$ & $\lambda_1$ & $\sum\limits_{k=1}^{N_u} \lambda_k$ \\
\midrule
\hline
{E01} &  254.7 & S,E & 0.14,0.69 & 0.13 & 0.57 & 2 & 4.53 & 0.0272 & 0.0517 \\
{E02} &  258.2 & S & 0.50 & 0.27 & 0.40  & 8 & 13.6 & 0.1508 & 0.2420 \\
{E03} &  258.4 & S,E & 0.24,0.39 & 0.19  & 0.35  & 7 & 13.0 & 0.1492 & 0.1922 \\
{E04} &  259.5 & S & 0.41 & 0.35 &  0.40  & 6 & 11.1 & 0.1422 & 0.2053 \\
{E05} &  261.1 & E & 0.82 & 0.72 &  0.67  & 7 & 14.4 & 0.1926 & 0.4916 \\
{E06} &  261.6 & S & 0.38 & 0.34 &  0.50  & 6 & 9.90 & $0.0521 + 0.0458$ & 0.1257 \\
{E07} &  263.4 & E & 0.69 & 0.61 &  0.53  & 8 & 15.7 & 0.1222 & 0.3868 \\
{E08} &  264.1 & S & 0.50 & 0.49 &  0.47  & 3 & 6.81 & $0.0344 + 0.0708i$ & 0.0828\\
{E09} &  267.0 & S & 0.60 & 0.46 &  0.48  & 6 & 17.5 & $0.1514 + 0.1097i$ & 0.6892\\
{E10} &  267.3 & S & 0.61 & 0.47 &  0.50  & 5 & 13.3 & $0.1074 + 0.0243i$ & 0.3991\\
{E11} &  267.4 & S,E & 0.42,0.48 & 0.41 &  0.47 & 5 & 12.2 & 0.0896 & 0.2216\\
{E12} &  267.6 & S & 0.70 & 0.44 &  0.49  & 3 & 10.7 & $0.0775 + 0.0390i$ & 0.1726\\
{E13} &  267.7 & S & 0.61 & 0.45 &  0.45  & 5 & 10.1 & $0.0231 + 0.1900i$ & 0.0914\\
{E14} &  267.9 & S & 0.42 & 0.39 &  0.50  & 6 & 14.0 & $0.0922 + 0.0312i$ & 0.3493\\
{E15} &  268.3 & E & 0.61 & 0.48 &  0.50  & 6 & 15.9 & $0.1911 + 0.0737i$ & 0.5934\\
{E16} &  268.7 & E & 0.63 & 0.53 &  0.37  & 5 & 10.9 & $0.0749 + 0.0775i$ & 0.2482\\
{E17} &  268.8 & E & 0.63 & 0.63 &  0.57  & 6 & 16.2 & $0.1249 + 0.1935i$ & 0.4468\\
{E18} &  269.2 & S,E & 0.49,0.54 & 0.49 & 0.46  & 6 & 16.9 & 0.1608 & 0.4230\\
{E19} &  270.7 & E & 0.64 & 0.61  & 0.49  & 8 & 17.0 & $0.1106 + 0.1261i$ & 0.4072\\
{E20} &  272.4 & S,E & 0.53,0.33 & 0.41 & 0.33 & 4 & 15.3 & $0.1014 + 0.1787i$ & 0.2977\\
{E21} &  273.3 & E & 0.49 & 0.48  & 0.46  & 6 & 19.6 & $0.1786 + 0.0852i$ & 0.5520\\
{E22} &  274.0 & E & 0.71 & 0.54  & 0.56 & 9 & 20.3 & 0.2440 & 0.7611\\
{E23} &  274.1 & S & 0.40 & 0.36 & 0.41  & 7 & 18.6 & $0.1318 + 0.1681i$ & 0.3942\\
{E24} &  275.7 & E & 0.61 & 0.56 & 0.49  & 7 & 16.0 & 0.1847 & 0.3497\\
{E25} &  275.8 & S,E & 0.54,0.53 & 0.38 & 0.40 & 8 & 17.2 & $0.1134 + 0.1611i$ & 0.4019\\
{E26} &  276.1 & S & 0.33 & 0.33 & 0.49 & 3 & 8.99 & $0.0284 + 0.1235i$ & 0.0752\\
{E27} &  277.2 & S & 0.43 & 0.40 & 0.48 & 4 & 8.56 & $0.0394 + 0.0896i$ & 0.1165\\
{E28} &  278.6 & E &  0.70 & 0.50 & 0.49 & 6 & 17.7 & $0.1037 + 0.0325i$ & 0.4475\\
{E29} &  278.8 & E & 0.48 & 0.49 & 0.44  & 5 & 14.6 & $0.1015 + 0.2180i$ & 0.3721\\
{E30} &  279.8 & E & 0.60 & 0.46 & 0.49  & 7 & 17.9 & 0.2433 & 0.4565\\
{E31} &  279.9 & E & 0.58 & 0.39 & 0.42  & 9 & 19.4 & 0.0723 & 0.3135\\
\hline
\bottomrule
\end{tabular}
\caption{\label{table:sol_summary}{Unstable equilibria (${\bf u}_0$) computed using initial conditions (IC) from simulation (S) and experiment (E), sorted {by} their $L2-$norm $\|{\bf u}_0\|$. $\min_{ic}D_0^{ic}$ is the {distance from an equilibrium to the nearest initial condition (experimental or numerical) that converged to it under Newton's iteration}.   {$\min_t D_0^{sim}(t)$  ($\min_t D_0^{exp}(t)$) are the distances of the} closest approach of turbulent trajectory in simulation (experiment) to {an} equilibrium. 
{$N_u$ is the number of unstable eigenvalues and $N_{KY}$ is the local Kaplan-Yorke dimension.
Eigenvalues are ordered by their real parts, with $\lambda_1$ being the most unstable one.}}}
\end{table*}

Among all of the equilibria calculated, six were found using both experimental and numerical initial conditions, resulting in a total of 31 distinct equilibria.
Due to the equivariance of the governing equation under $\rsym$, if ${\bf u}_0$ is an equilibrium which is not invariant under $\rsym$, then its rotated copy {$\rsym{\bf u}_0\,(\neq{\bf u}_0)$} is also an equilibrium. 
Consequently, distinct initial conditions may converge to either ${\bf u}_0$ or {$\rsym{\bf u}_0$}.
Hence, the converged equilibria were tested for presence of symmetry related copies to determine the total number of distinct ones.
Table \ref{table:sol_summary}  lists all the distinct equilibria we computed, labeled in ascending order of their $L_2$-norm $\|{\bf u}_0\|$. 
Also listed are the data sets which contained the initial conditions that converged to a particular solution: simulation (S), experiment (E), or both (S,E). 

The Newton-Krylov solver initialized using ${\bf u}_{ic}$ can, in principle, converge to an equilibrium ${\bf u}_0$ that lies far away from that initial condition. 
The degree of similarity between these two states can be quantified using the normalized distance
\begin{equation}\label{eq:eq_fp2ic}
D_0^{ic} =  \frac{\|{\bf u}_{ic}-{\bf u}_0\|}{\|{\bf u}_0\|}.
\end{equation}
and is listed in Table \ref{table:sol_summary} for each equilibrium.
When several different initial conditions converged to an equilibrium, the smallest $D_0^{ic}$ was used.
To relate the magnitude of $D_0^{ic}$ with the visual similarity between the flow fields ${\bf u}_{ic}$ and ${\bf u}_0$ in the physical space, we have included in the supplemental material the vorticity fields of all the equilibria and the nearest initial conditions.
A quick comparison suggest that ${\bf u}_{ic}$ and ${\bf u}_0$ appear similar when $D_0\lesssim 0.60$.
For example, the converged solution E10 and the corresponding initial condition shown in \reffig{sample_sols_sim}(b) bear a striking resemblance despite differing by $D_0^{ic} = 0.61$.

To test how close the turbulent flows in experiments and simulations approach each equilibrium, we computed the minimal (normalized) distance
\begin{equation}\label{eq:eq_fp2turb}
D_0(t) =  \min_g\frac{\|{\bf u}(t)-g{\bf u}_0\|}{\|{\bf u}_0\|}.
\end{equation}
Table \ref{table:sol_summary} lists the {minimal} distances $\min_tD^{sim}_0(t)$ and $\min_tD^{exp}_0(t)$  from each equilibrium to turbulent trajectories in simulation and experiment, respectively.  
The data presented in Table \ref{table:sol_summary} shows that, based on a threshold of 0.6, 
all but four (E05, E07, E17, and E19) equilibria were visited by the turbulent flow in {\it both} experiment and simulations, validating their dynamical relevance.
Moreover, despite a shorter time series in the experiment, $\min_t D_0^{exp}(t)$ is not systematically higher than $\min_t D_0^{sim}(t)$ (except for E01,), confirming that equilibria of the 2D model are indeed dynamically relevant in the experiment.  
We note that only four equilibria computed -- E01, E05, E06, and E26 -- 
are invariant under $\rsym$ and trajectories in the experiment, where $\rsym$ is weakly broken, may not approach these solutions as closely as numerical trajectories might.

\subsection{Invariant Manifolds of Equilibria}\label{sec:manifolds}

The dynamical role of equilibria goes beyond causing slowdowns in the evolution of turbulent trajectories entering their neighborhoods. 
Their stable and unstable manifolds shape the state space geometry, guiding nearby turbulent trajectories, which follow the stable manifold of an equilibrium on approach and the unstable manifold on departure.
The local orientation and the dimensionality of the stable and unstable manifolds are determined, respectively, by the stable and unstable eigenvectors of the corresponding equilibrium.
{As Table \ref{table:sol_summary} shows, the number of unstable directions $N_u$ is small: it} varies between 2 and 9 for all the equilibria we identified, which is a tiny fraction of the dimensionality of the full state space, $N_f\approx 2\times10^5$.

On the contrary, the total number of stable directions, $N_f-N_u$, is very large, which appears to suggest a dramatic asymmetry between stable and unstable manifolds.
However, only a tiny fraction of the {degrees of freedom associated with the stable manifold} play a role in sustained fluid turbulence, 
{with dissipation constraining the dynamics to} a relatively low-dimensional chaotic attractor  \cite{hopf_1948, ruelle_1971, brandstater_1983}.   
The number of dynamically relevant {degrees of freedom} {in the neighborhood of an equilibrium} can be estimated  by computing the local Kaplan-Yorke dimension \cite{farmer_1983}
\begin{equation}\label{eq:kydim}
N_{KY} = k_0 + {\frac{1}{|\Re(\lambda_{k_0+1})|}}\sum\limits_{k = 1}^{k_0} \Re(\lambda_{k}),
\end{equation}
where the eigenvalues $\lambda_k$ are sorted by their real parts $\Re(\lambda_k)$ in descending order and $k_0$ is the largest integer for which the sum on the right-hand-side of \eqref{eq:kydim} is non-negative.
For the equilibria considered here $N_{KY} (\approx 2N_u)$ varies roughly between 4 and 20, suggesting that the number of dynamically relevant stable directions $N_{KY}-N_u$ is $O(10)$. {To test if $N_{KY}$ corresponding to dynamically relevant ECSs is comparable to that of the attractor, we computed the Lyapunov spectrum of the chaotic attractor using continuous Gram-Schmidt orthogonalization \cite{christiansen_1997a}. Temporal average of the  spectrum showed $N_{KY}\approx 15$, which  is indeed  comparable to $N_{KY}$ corresponding to the equilibria we computed.}  

Although the numbers of dynamically relevant stable and unstable directions in the neighborhood of the equilibria listed in Table \ref{table:sol_summary} are comparable, in the remainder of this paper we will focus on unstable manifolds.
They can be computed more easily using forward time integration and 
are also more useful in practice, e.g., allowing forecasting of the evolution of turbulent flows \cite{suri_2017a}. 
Two different examples, one with a pair of unstable eigenvalues with comparable magnitude and another characterized by a dominant real 
eigenvalue, will be used below to illustrate the dynamical role of the manifolds.

\subsubsection{Two-Dimensional Unstable Manifold}\label{sec:2dmanifold}

Equilibrium E01 shown in \reffig{sample_sols_sim}(a) has just two unstable eigenvectors $\uv{e}_1$ and $\uv{e}_2$, both with real eigenvalues $\lambda_{1} = 0.0272$ and $\lambda_{2} = 0.0245$. 
The perturbations corresponding to $\uv{e}_1$ and $\uv{e}_2$ in physical space are included in Appendix \ref{sec:projections}.
The 2D unstable manifold of E01 is locally tangent to the plane defined by $\uv{e}_1$ and $\uv{e}_2$, but deviates from this plane farther away due to nonlinearity. 
It was therefore computed using a dense set of 1440 trajectories ${\bf u}_{\theta}(t')$ with initial conditions uniformly distributed around a circle ${\bf u}_{\theta}(0) = {\bf u}_0 +\epsilon \cos(\theta)\uv{e}_1 + \epsilon\sin(\theta)\uv{e}_2$ with $\epsilon = 10^{-4}\times\|{\bf u}_0\|$.
The time $t'$ and angle $\theta$ uniquely parametrize this 2D manifold.

\begin{figure}[!t]
\centering
{\includegraphics[width=3.3in]{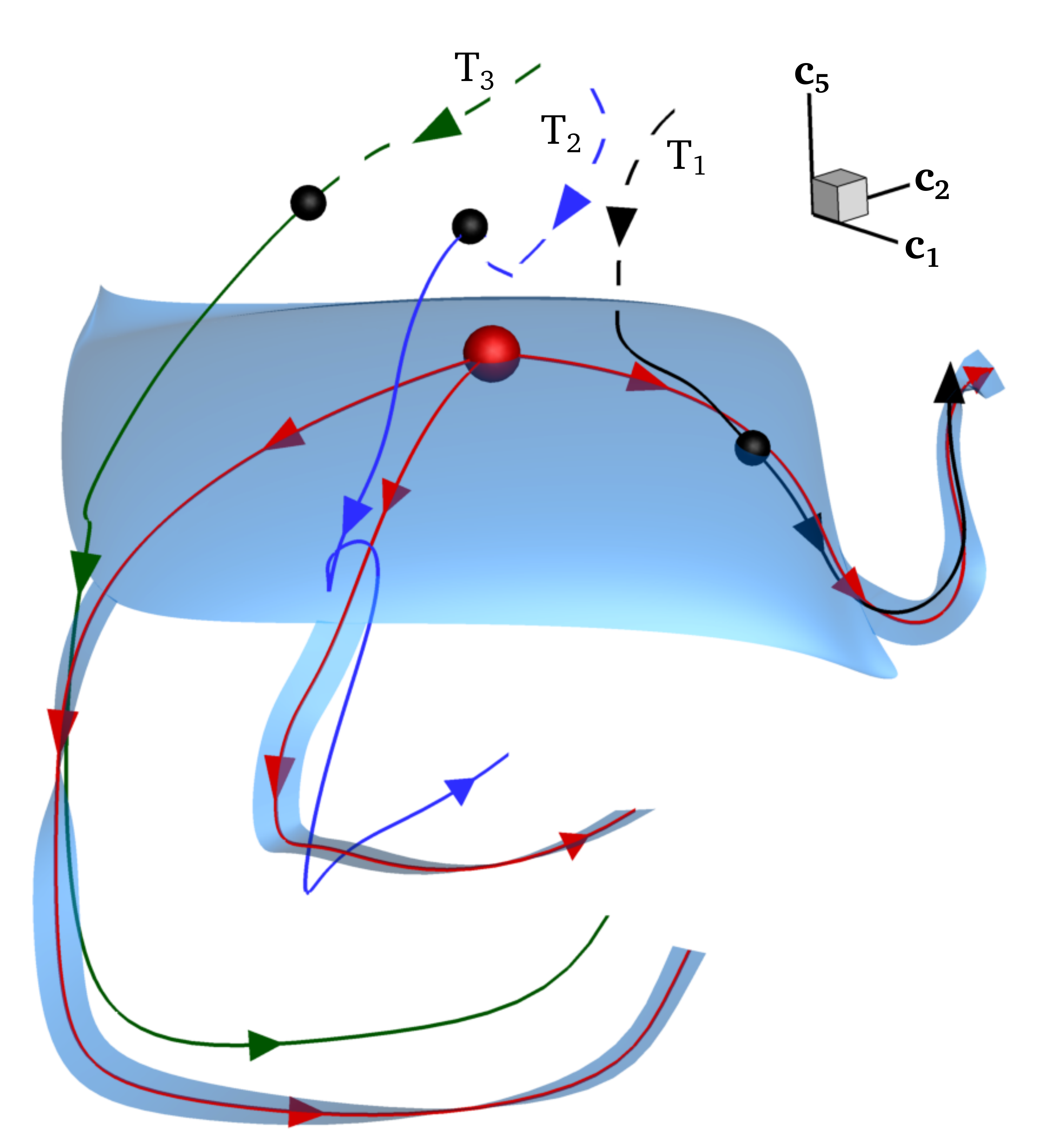}}
\caption{\label{fig:manifold_2d} 
{Turbulent trajectories ($\textrm{T}_1$, $\textrm{T}_2$, $\textrm{T}_3$) from numerical simulation {shadowing} the 2D unstable manifold (blue surface) of equilibrium E01 (red sphere). Black spheres on $\textrm{T}_1$, $\textrm{T}_2$, $\textrm{T}_3$ mark minima in $s(t)$.
The portion of the unstable manifold shown was constructed using a one-parameter family of trajectories. 
Blue ribbons and red curves represent manifold extensions and reference manifold trajectories that $\textrm{T}_1$, $\textrm{T}_2$, and $\textrm{T}_3$ follow. 
The procedure used to construct this low-dimensional projection and the definition of basis vectors ${\bf c}_i$ are described in Appendix \ref{sec:projections}.}} 
\end{figure}

A few of the manifold trajectories ${\bf u}_{\theta}(t')$ along with a portion of the unstable manifold are shown in figure \ref{fig:manifold_2d}, which was constructed by projecting the high-dimensional state space onto an orthonormal basis spanned by the unstable eigenvectors $\uv{e}_1$, $\uv{e}_2$, and a stable eigenvector $\uv{e}_5$ {(cf. Appendix \ref{sec:projections})}.
Note that all the trajectories lying in the manifold exhibit very little curvature near the equilibrium, which is a consequence of the two unstable eigenvalues being nearly equal.
However, away from E01, the nonlinearity of the governing equation results in significant curvature of the manifold and nearby trajectories.

As we mentioned previously, sufficiently close passes to E01 were not observed in the experiment. In contrast, turbulent trajectories  ${\bf u}(t)$ in the simulation were found {to approach} E01 very closely on numerous occasions.  
{Three such trajectories with $\min_t D_0$ = 0.13 ($\textrm{T}_1$), 0.26 ($\textrm{T}_2$), and 0.26 ($\textrm{T}_3$) are shown in \reffig{manifold_2d}.}
The segments of these trajectories shown are approximately $10\tau_c$, $8\tau_c$, and $7\tau_c$-long, respectively. The trajectory $\textrm{T}_1$ corresponds to the deepest minimum of $s(t)$ as well as the closest approach to E01 across the entire time series (cf. \reffig{sss_sim}).

Figure \ref{fig:manifold_2d} shows that nearby turbulent trajectories indeed approach E01 along the stable manifold (in this projection along {\bf c${_5}$}) and subsequently depart following its unstable manifold.  
The segment of each turbulent trajectory as it approaches E01 is plotted using a dashed curve, while that following the unstable manifold  is plotted using a solid curve.  
Notice that turbulent trajectories departing the neighborhood of E01 are guided by the unstable  manifold even very far away from E01, where the linearization used to compute the eigenvalues and eigenvectors completely breaks down, as illustrated by the curvature of the manifold.

To make sure that the low-dimensional projection accurately reflects the dynamics in the full state space, we performed a quantitative analysis of the turbulent trajectories passing through the neighborhood of E01. 
{We started by computing the instantaneous distance $D_1$ between a turbulent trajectory ${\bf u}(t)$ and each manifold trajectory ${\bf u}_\theta(t')$:
\begin{align}\label{eq:eq_1dtoturb}
D_1(t,\theta) = \min_{t'}\frac{\|{\bf u}(t)-{\bf u}_{\theta}(t')\|}{\|{\bf u}(t)\|}.
\end{align}
For a point ${\bf u}(t)$ on the turbulent trajectory, $D_1$ is the distance to the closest point -- parametrized by $t'$ -- on a manifold trajectory ${\bf u}_\theta(t')$.
We then computed the instantaneous distance from the turbulent trajectory to the entire 2D unstable manifold parametrized by $\theta$ and $t'$:
\begin{align}\label{eq:eq_2dtoturb}
D_2(t) =  \min_{\theta}D_1(t,\theta) = \min_{\theta, t'}\frac{\|{\bf u}(t)-{\bf u}_{\theta}(t')\|}{\|{\bf u}(t)\|}.
\end{align}
$D_2$ can be used to identify the manifold trajectory ${\bf u}_\theta$ which is the closest to the turbulent trajectory ${\bf u}(t)$ at a given instant.}

\begin{figure}[!t]
\centering
\qquad \includegraphics[width=3.1in]{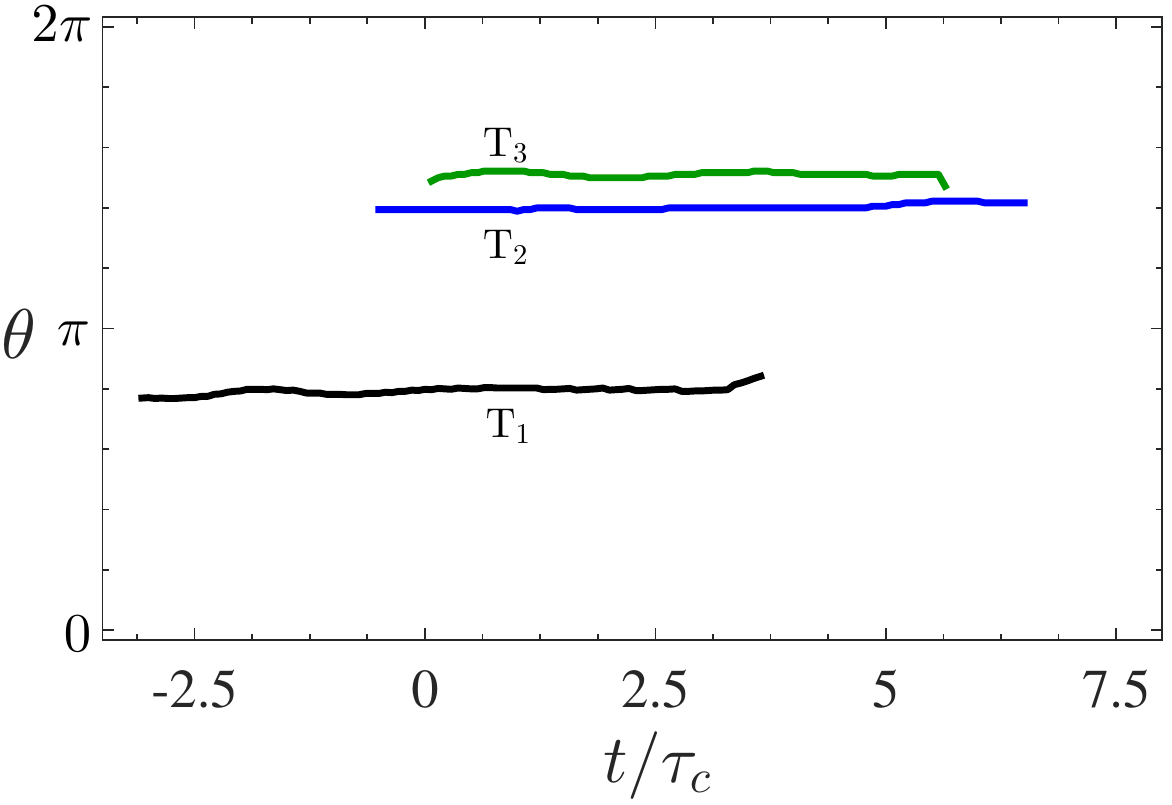}
\caption{\label{fig:2d_theta}  The angle $\theta$ for the manifold trajectory which is closest to the points ${\bf u}(t)$ on a nearby turbulent trajectory.} 
\end{figure}

Figure \ref{fig:2d_theta} shows the value of $\theta$ that corresponds to $D_2(t)$ for the three turbulent trajectories that appear in \reffig{manifold_2d}. 
In each case, $t=0$ corresponds to a minimum of $s(t)$, and only the temporal interval during which $\theta$ does not experience abrupt changes is plotted. 
We find that there is a fairly long time interval centered around $t=0$ over which $\theta\approx \theta_c =\theta|_{t=0}$ is essentially constant, i.e, each turbulent trajectory follows a specific reference trajectory ${\bf u}_{\theta_c}(t')$ lying in the manifold. The reference trajectories in the unstable manifold corresponding to $\textrm{T}_1$, $\textrm{T}_2$, and $\textrm{T}_3$ are plotted in red in \reffig{manifold_2d}.

\begin{figure}[!t]
\centering
\includegraphics[width=3.3in]{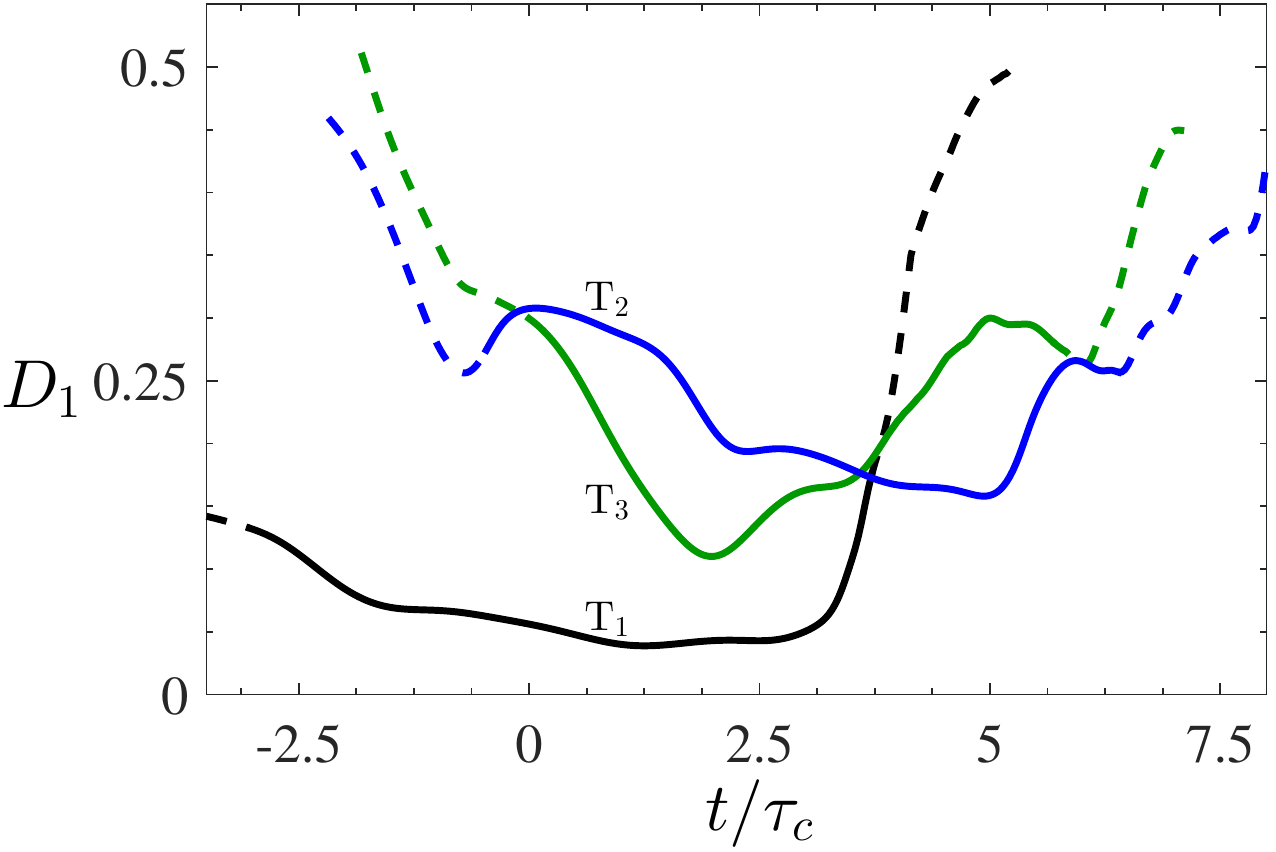}
\caption{\label{fig:2d_follow} {The distance between the turbulent trajectory ${\bf u}(t)$ and the corresponding manifold trajectory ${\bf u}_{\theta_c}(t')$ (see text).}} 
\end{figure}

The distance between each turbulent trajectory and the corresponding manifold reference trajectory is given by {$D_1(t,\theta_c)$.
In figure \ref{fig:2d_follow} we plot $D_1(t,\theta_c)$ for each of the three turbulent trajectories shown in \reffig{manifold_2d}. 
Solid curves indicate the time intervals over which $\theta\approx\theta_c$ and so $D_1(t,\theta_c)\approx D_2(t)$} is an accurate estimate of the distance from the turbulent trajectory to the entire unstable manifold. 
The relatively low values of the distance indicate that turbulent trajectories follow the corresponding manifold trajectories quite closely in the full state space for at least $3.5\tau_c$ following the instant at which $s(t)$ is a minimum.

\begin{figure}[!t]
\centering
\subfloat[]{\includegraphics[height=2in]{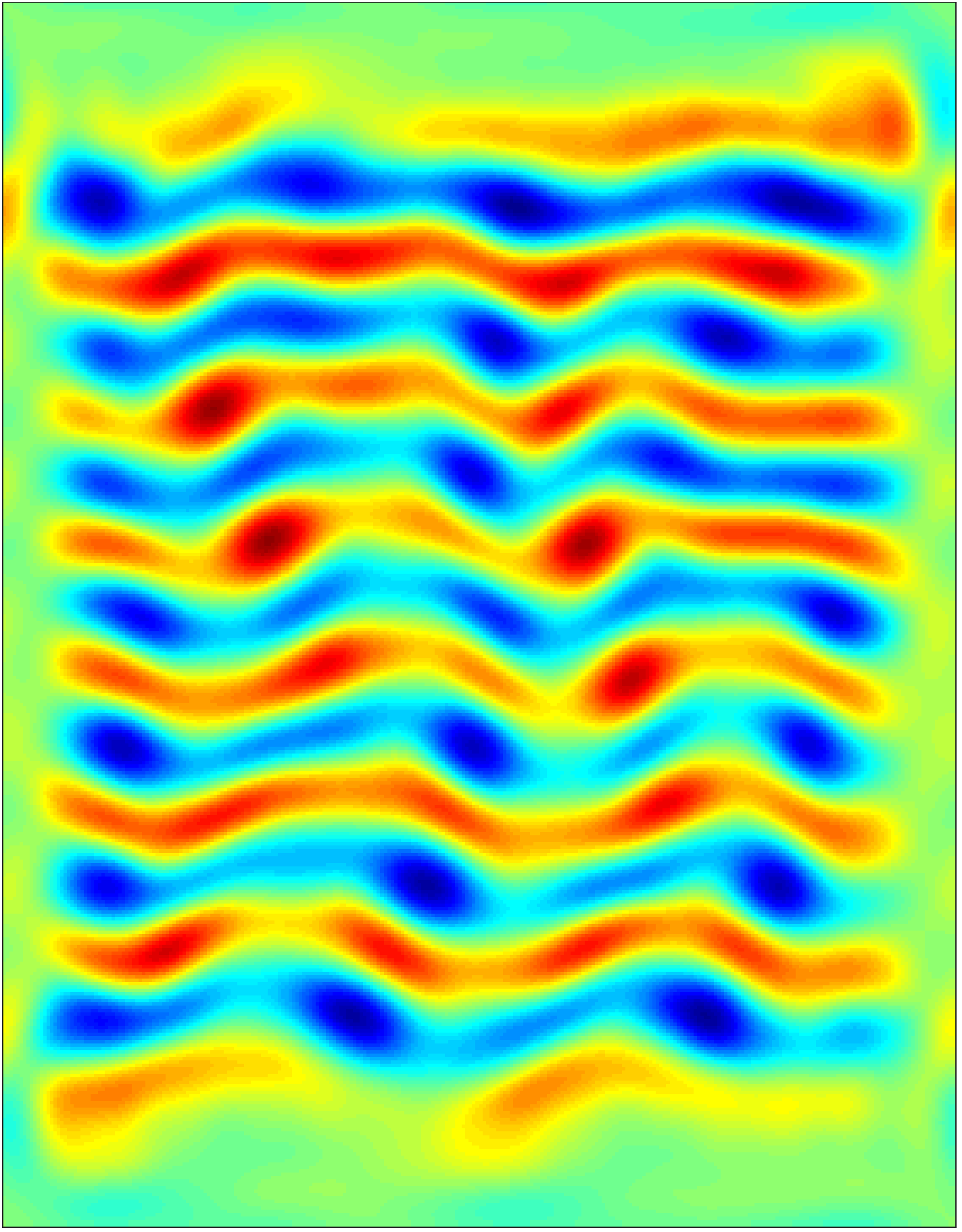}} \hspace{1.75mm} 
\subfloat[]{\includegraphics[height=2in]{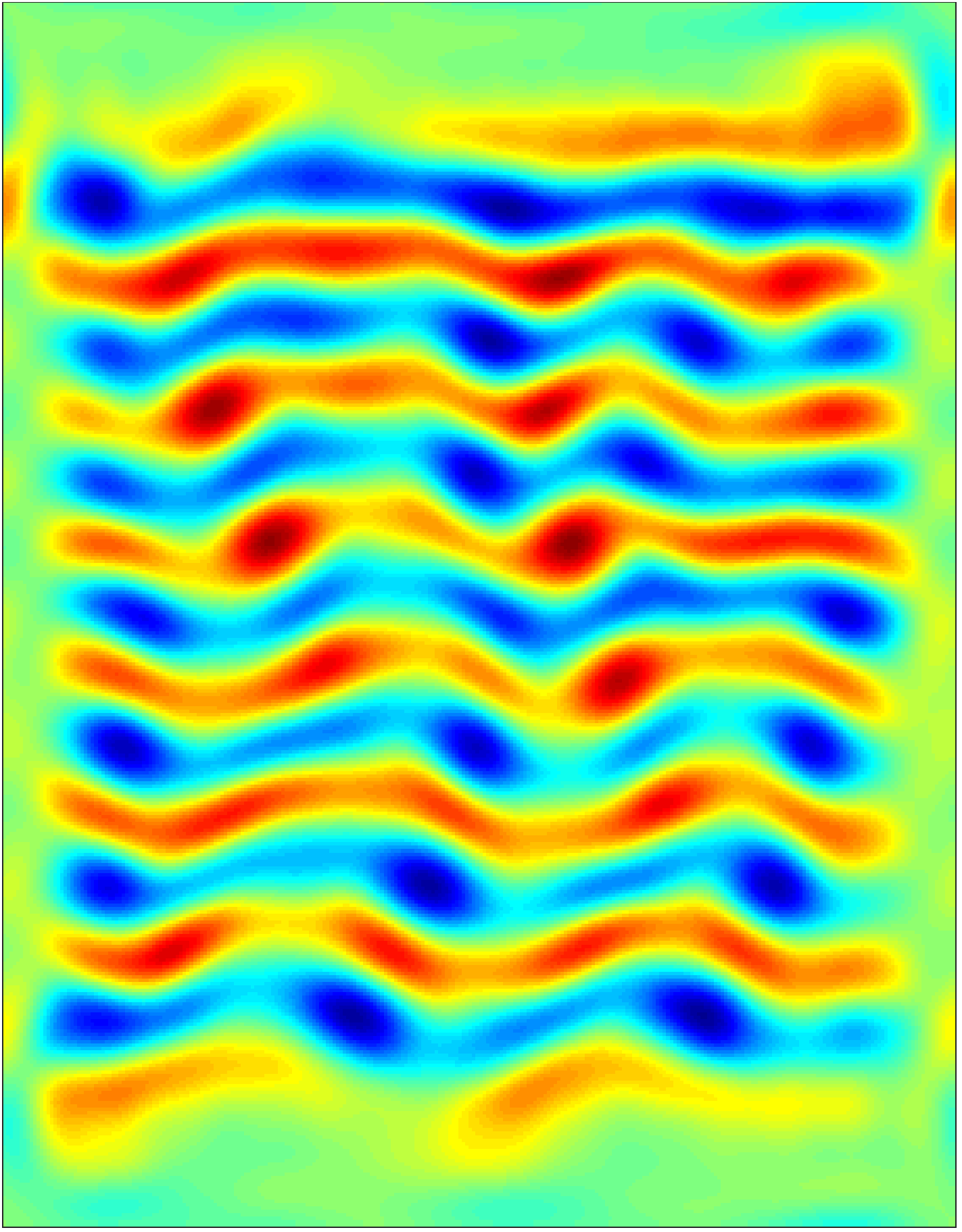}}
\caption{\label{fig:manifold_2d_flows} {Flow fields for (a) the turbulent trajectory $\textrm{T}_1$ in \reffig{manifold_2d} at $t/\tau_c= 3.75$ and (b) the nearest point on the reference manifold trajectory.}
} 
\end{figure}

Additional evidence for the dynamical role of the unstable manifold is provided in figure \ref{fig:manifold_2d_flows}, which compares the flow field {on ${\textrm T}_1$} -- at a location which is far from E01 in the full state space -- with the nearest point on the reference manifold trajectory.  
These states correspond to the ends of the respective trajectories in \reffig{manifold_2d}, where $D_1\approx 0.3$.
The striking similarity between these flow fields once again confirms the hypothesis that turbulent flow is guided in state space by the unstable manifold (of E01 in this case) even when the flow has evolved substantially far away from the unstable equilibrium. 

As \reffig{2d_follow} shows, all three turbulent trajectories continue to approach the unstable manifold as they evolve farther away from the equilibrium. 
Indeed, the unstable manifold should be locally attracting for all initial conditions in the immediate neighborhood of the corresponding equilibrium. 
The unusual aspect here is that none of the three turbulent trajectories come particularly close to E01{: the separations $D_2(0)\approx D_1(0)$ are} between 0.1 and 0.3, which is outside the linear neighborhood of E01.
Since the flow is chaotic, the unstable manifold becomes repelling far from the equilibrium. So even turbulent trajectories that approach the unstable manifold fairly closely eventually diverge away from it (cf. \reffig{2d_follow}).
This divergence imposes an inherent limit {on} the distance in state space over which  turbulent trajectories are guided by any of the unstable manifolds.

\begin{figure}[!t]
\centering
\includegraphics[width=3.3in]{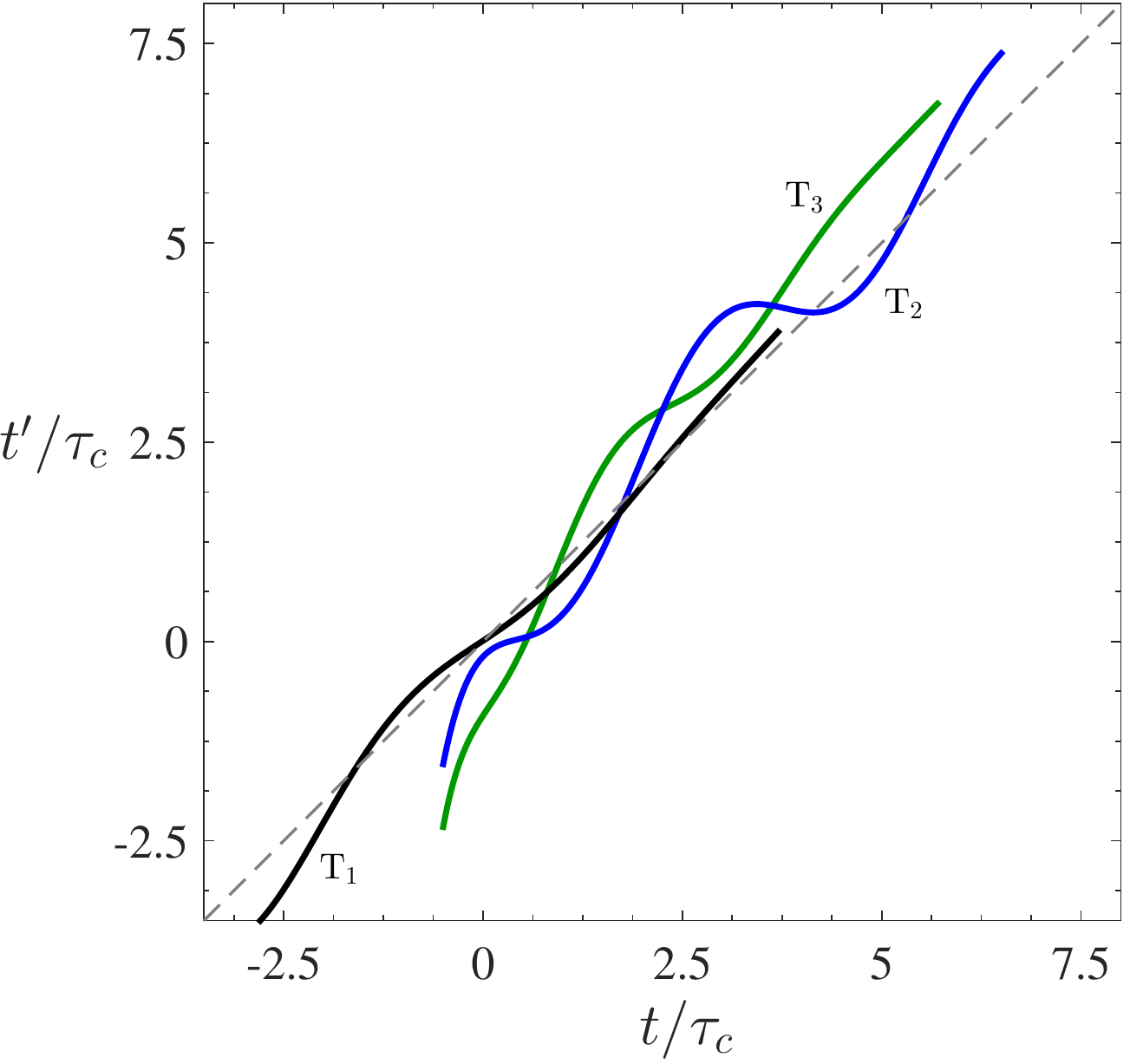}
\caption{\label{fig:2d_sync} {Comparison between} the rates of evolution along manifold and turbulent trajectories.}
\end{figure}

In conclusion of this section, we comment on a dynamical aspect of the problem.  
The evidence we  presented illustrates the geometrical role of the unstable manifold of E01 and particular manifold trajectories.
However, one can ask whether the rate of evolution along the turbulent trajectory ${\bf u}(t)$ is the same as that for the corresponding manifold trajectory ${\bf u}_{\theta_c}(t')$.
Figure \ref{fig:2d_sync} shows the evolution of the ``manifold time'' $t'$, which corresponds to points along ${\bf u}_{\theta_c}$ closest to ${\bf u}(t)$.
Note that the origin for $t'$ is arbitrary and was chosen to minimize the difference between $t$ and $t'$.
Although, for all three turbulent trajectories, the graphs of $t'(t)$ cluster around the diagonal, which corresponds to identical rates of evolution for $t$ and $t'$, there are notable deviations.
For instance, we find a rather unexpected feature in the evolution of the turbulent trajectory $\textrm{T}_2$: $t'$ {\it decreases} for $3\lesssim t/\tau_c\lesssim 4$.
Indeed, during the corresponding time interval, this turbulent trajectory makes a small loop in figure \ref{fig:manifold_2d}.
This odd behavior has to do with weakly stable degrees of freedom, which feature oscillatory dynamics.

\subsubsection{Seven-Dimensional Unstable Manifold}\label{sec:1dsubmanifold}

\begin{figure}[!t]
\centering
\subfloat[]{\includegraphics[height=2in]{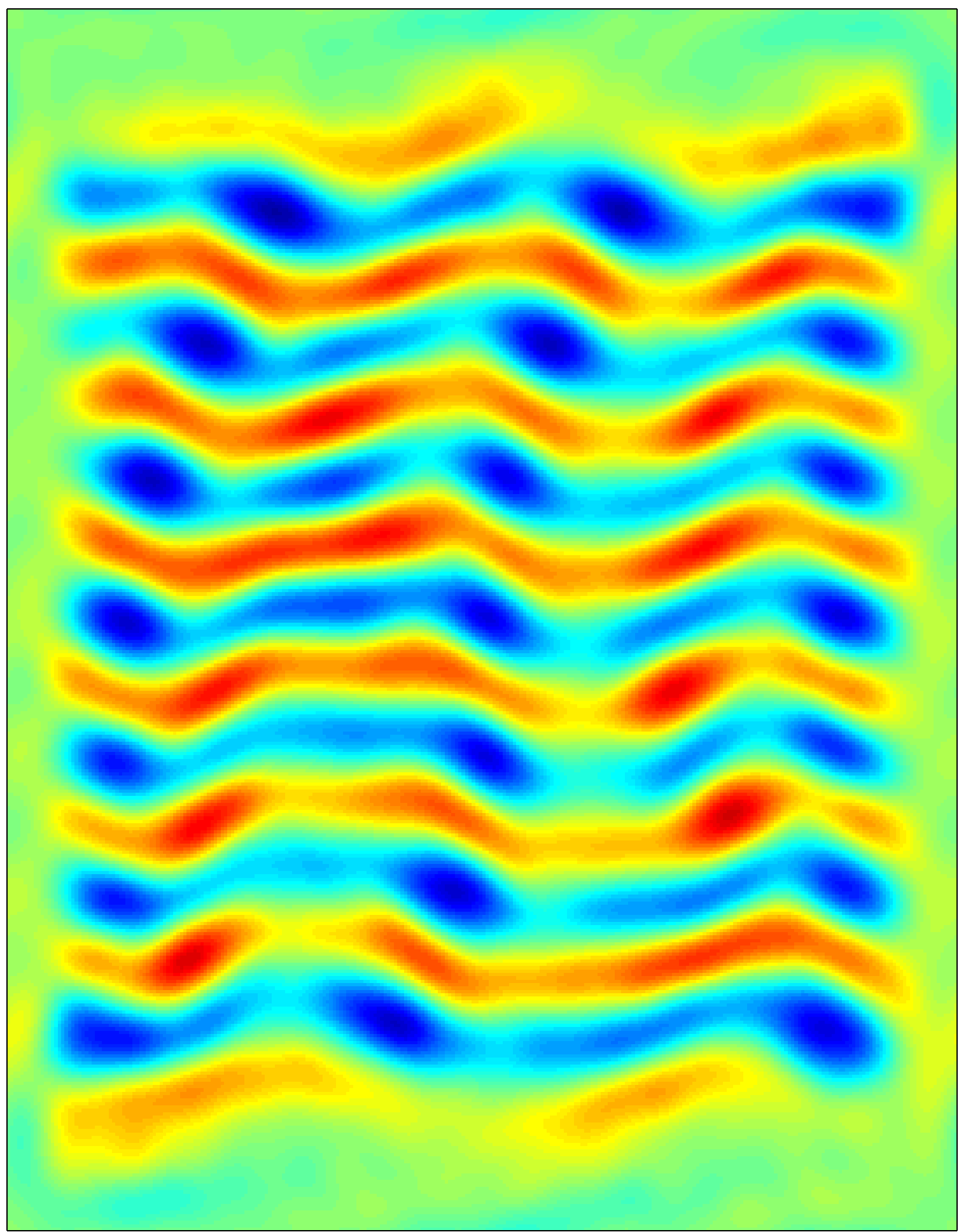}}  \hspace{1.85mm}
\subfloat[]{\includegraphics[height=2in,angle=0]{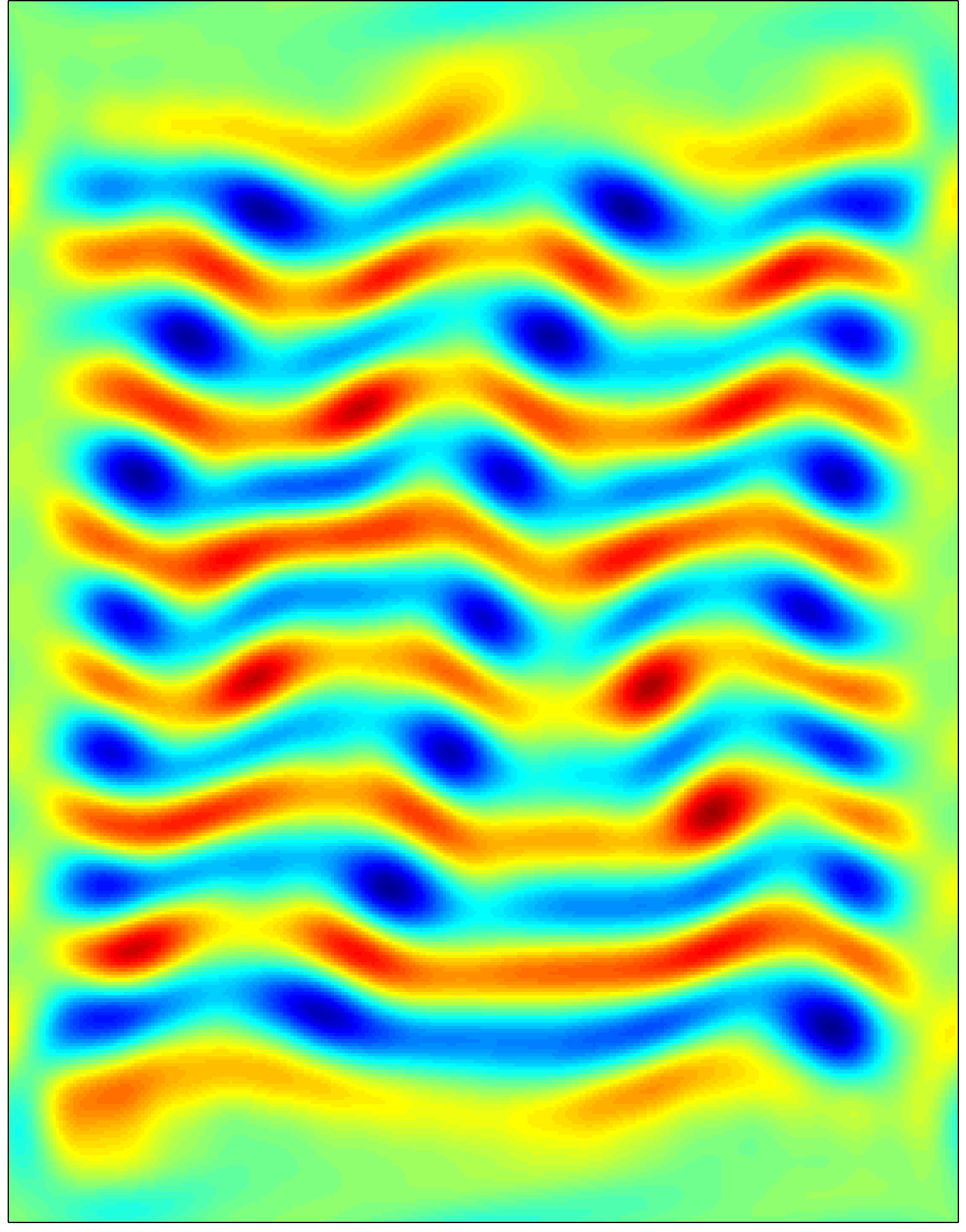}}
\caption{\label{fig:ecs_q1dm} {(a) A flow field from the experiment during a close pass to (b) equilibrium E03 which has a seven-dimensional unstable manifold.} 
}
\end{figure}

The analysis presented in the previous section suggests that turbulent evolution, following a deep minimum in $s(t)$, can be forecast for a few correlation times by constructing the unstable manifold of the nearby equilibrium. 
Such forecasting in both experiment and simulation for a close pass to E03 (cf. \reffig{ecs_q1dm}) was previously reported by Suri \etal \cite{suri_2017a}. 
Here, we extend that study by providing quantitative estimates for the separation between the unstable manifold of E03 and nearby turbulent trajectories.

\begin{figure}[!htbp]
\begin{center}
{\includegraphics[width=3.3in]{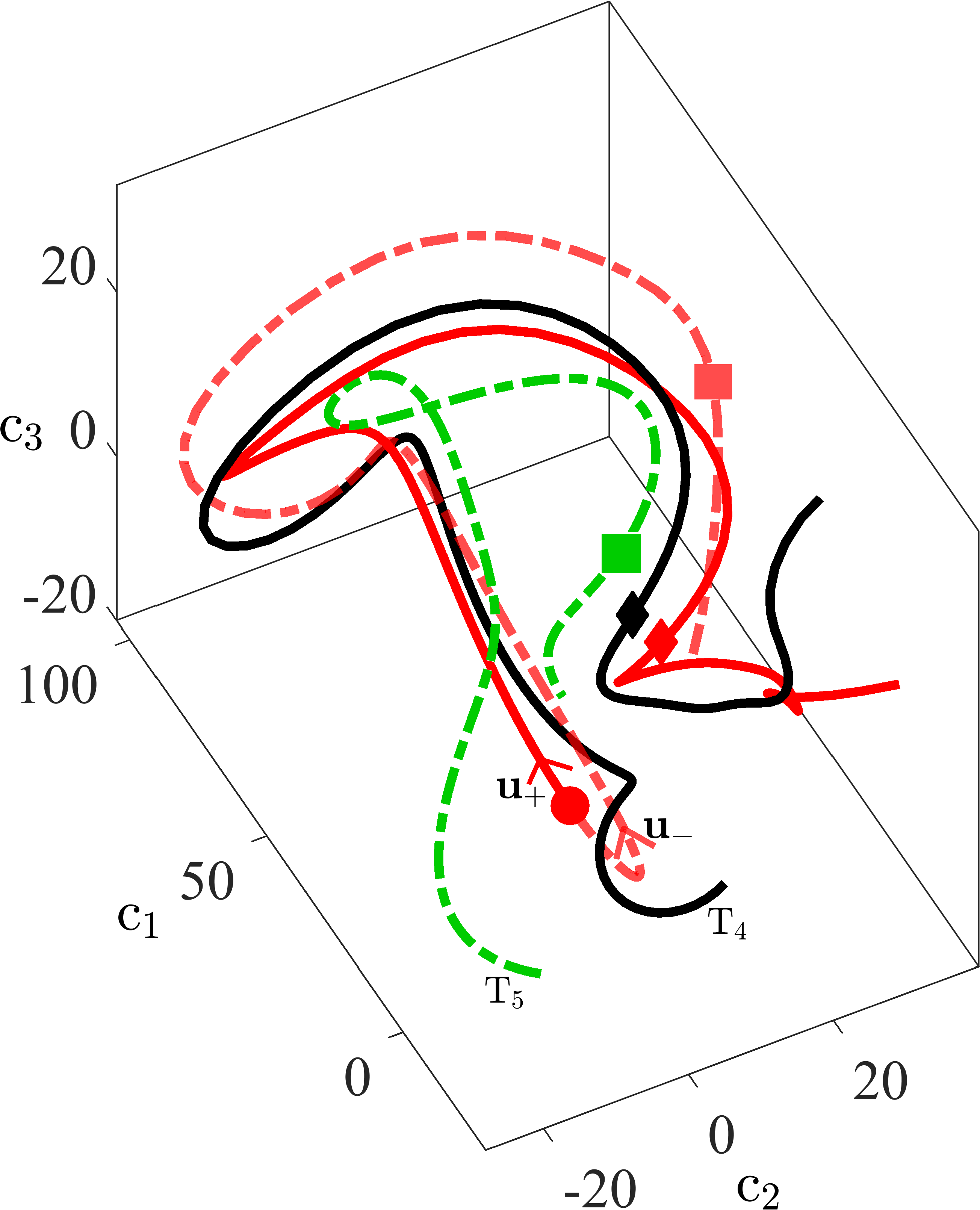}}
\end{center}
\caption{\label{fig:manifold_1d} The dominant unstable submanifold ${\bf u}_+(t')$  (solid red) and ${\bf u}_-(t')$ (dashed red) of equilibrium E03 (red sphere). {Also shown are turbulent trajectories from  simulation ($\textrm{T}_4$ in black) and experiment ($\textrm{T}_5$ in green) that shadow ${\bf u}_+(t')$ and ${\bf u}_-(t')$, respectively.} Details of the projection coordinates are provided in Appendix \ref{sec:projections}.} 
\end{figure}

Equilibrium E03 has a seven-dimensional unstable manifold with a leading real eigenvalue $\lambda_1$ = 0.1492 and three pairs of  unstable complex conjugate eigenvalues $\lambda_{2,3} = 0.0147\pm0.1680i$, $\lambda_{4,5} = 0.0045\pm  0.1104i$, and $\lambda_{6,7} = 0.0009\pm 0.4500i$. 
Due to its relatively high dimensionality, constructing the corresponding unstable manifold is exceedingly data intensive.
However, the spectral gap between the eigenvalues implies that not all unstable degrees of freedom are equally important dynamically \cite{gibson_2008,halcrow_2009}.
Trajectories starting from generic initial conditions close to E03 should quickly align in the direction of the leading eigenvector $\pm\uv{e}_1$ (the corresponding perturbation in the physical space is shown in Appendix \ref{sec:projections}). 

The strong focusing effect implies that the dynamically relevant portion of the unstable manifold of E03 is effectively one-dimensional (1D) and is shaped by the pair of trajectories ${\bf u}_{\pm}(t')$ departing from E03 along $\pm\uv{e}_1$.
The union of these two trajectories, which will be referred to as the dominant unstable submanifold in the following discussion, is shown as solid and dashed red curves in figure \ref{fig:manifold_1d}.
Also shown are E03 (red sphere) and turbulent trajectories from simulation {($\textrm{T}_4$)} and experiment {($\textrm{T}_5$)} in its neighborhood.
The figure is a  projection of the state space onto an orthogonal basis constructed using $\uv{e}_1$ and the eigenvectors $\uv{e}_6$ and $\uv{e}_7$ associated with the complex conjugate eigenvalue pair $\lambda_{6,7}$ (cf. Appendix \ref{sec:projections}).  
It is interesting to point out that ${\bf u}_{+}(t')$ appears nearly straight even well outside of the linear neighborhood of E03, while ${\bf u}_{-}(t')$, which is initially oriented in the opposite direction, turns around not far from the equilibrium and starts following ${\bf u}_{+}(t')$.
This somewhat odd observation is not an artifact of the projection, as confirmed by the evolution of the corresponding flow fields in the physical space, which appear quite similar for an extended period of time.

\begin{figure}[!t]
\begin{center}
{\includegraphics[width=3.3in]{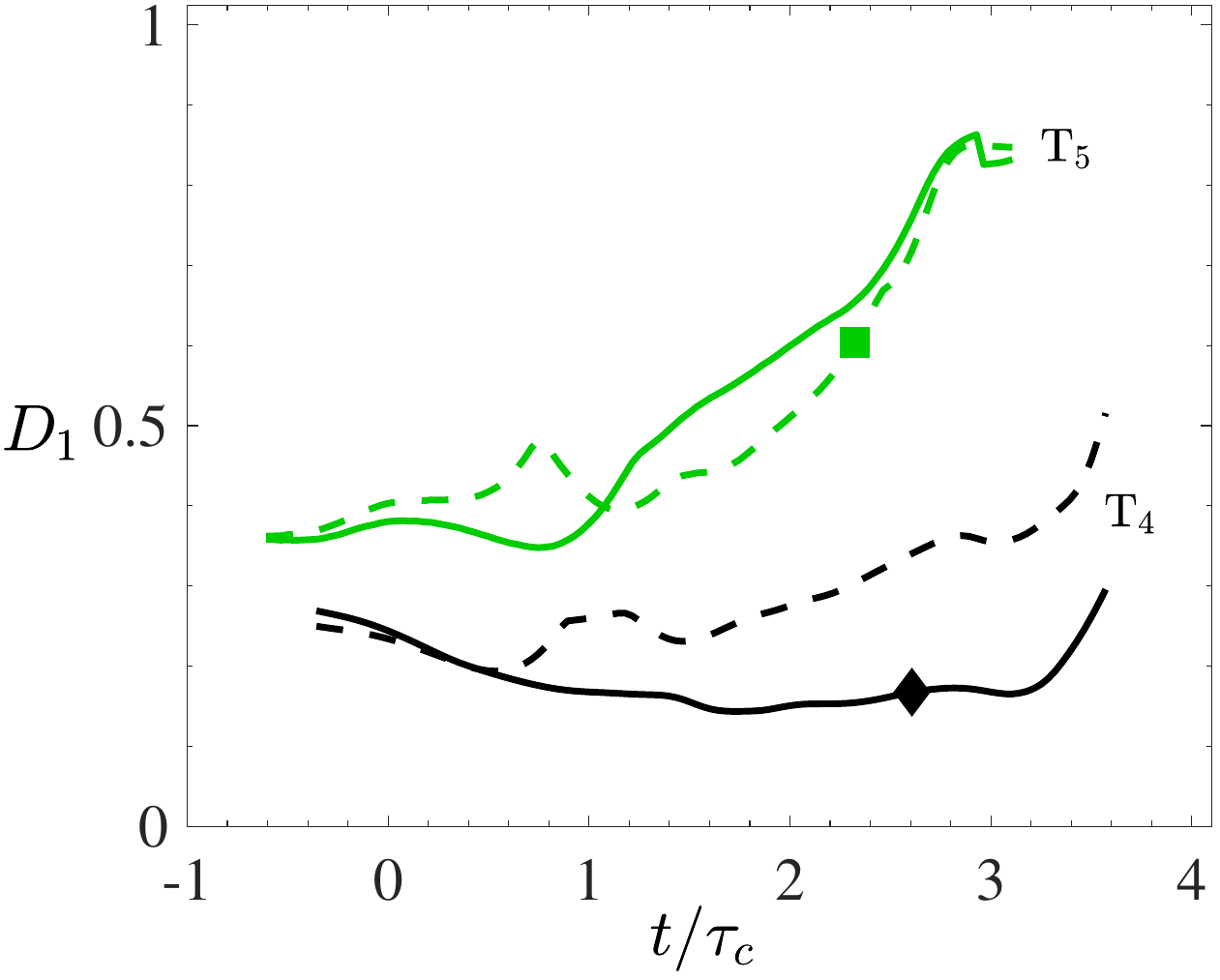}}
\end{center}
\caption{\label{fig:1d_follow} {Distance $D_1(t)$ from turbulent trajectories ${\bf u}(t)$ in simulation ($\textrm{T}_4$) and experiment ($\textrm{T}_5$) to the 1D submanifold. Solid (dashed) curves correspond to distance from ${\bf u}_+(t')$ (${\bf u}_-(t')$). The diamond and square highlight separation between states indicated using the same symbols in \reffig{manifold_1d}}.} 
\end{figure} 
\begin{figure}[!h]
\centering
\subfloat[]{\includegraphics[height=2in]{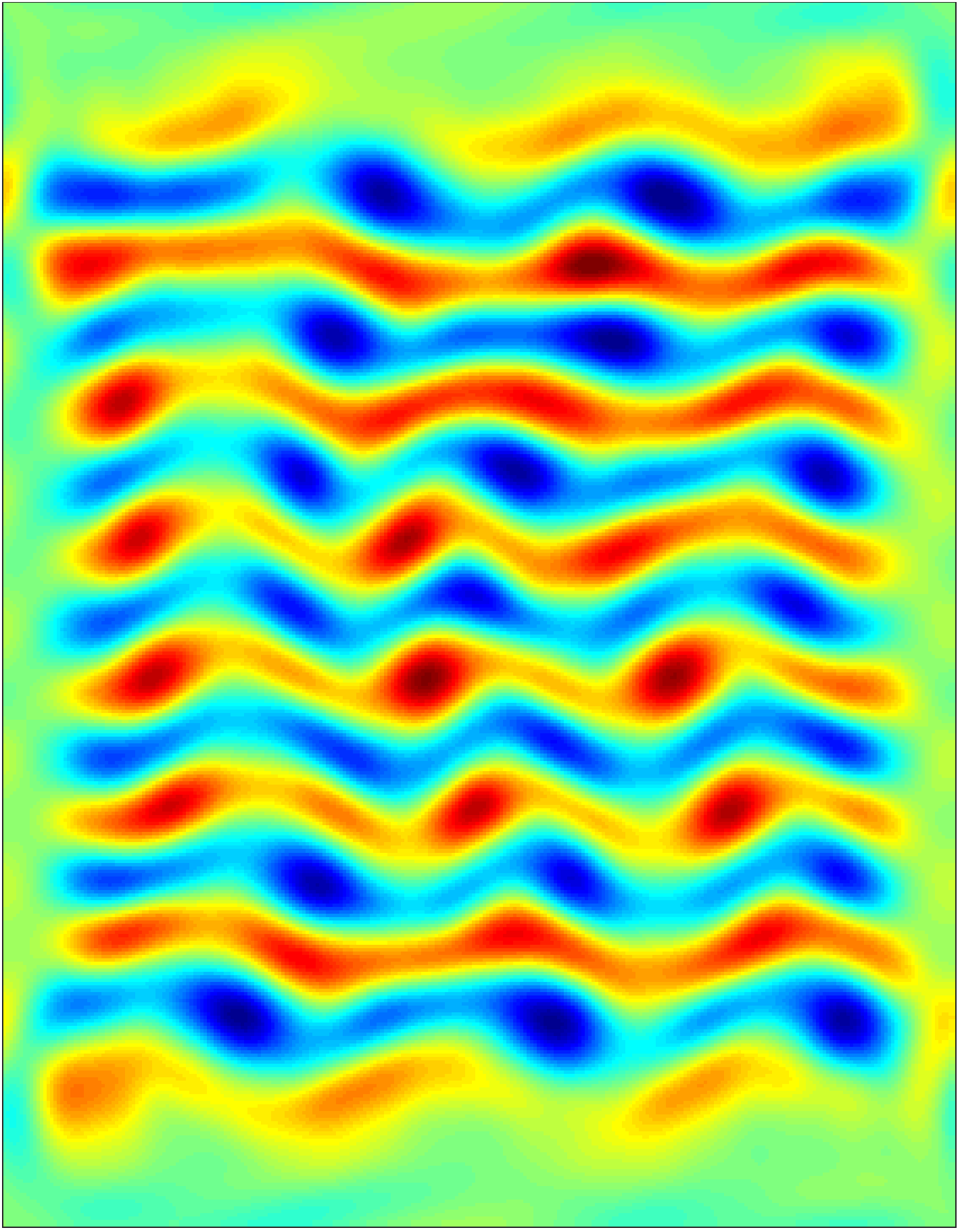}} \hspace{1.75mm} 
\subfloat[]{\includegraphics[height=2in]{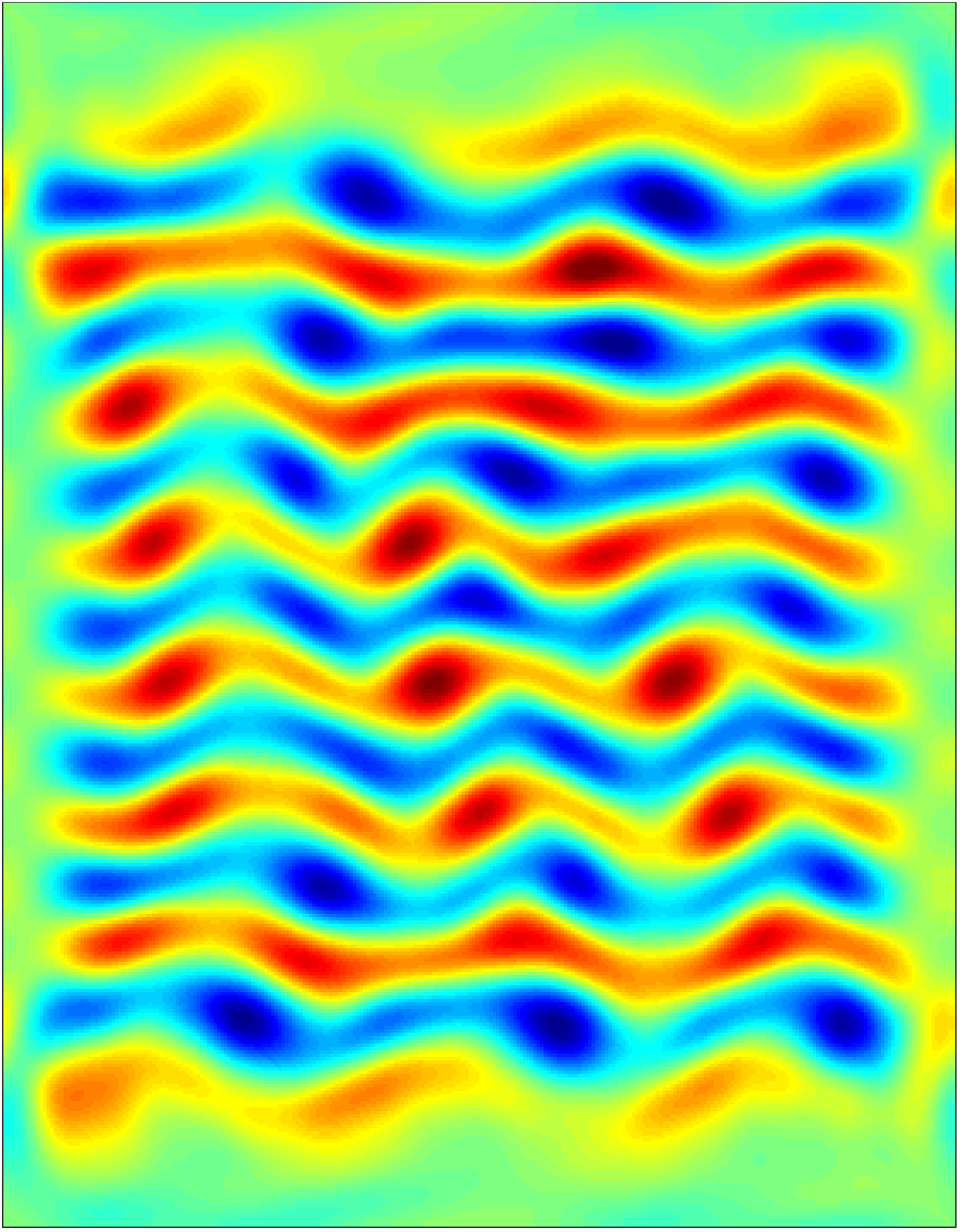}}\\
\subfloat[]{\includegraphics[height=2in]{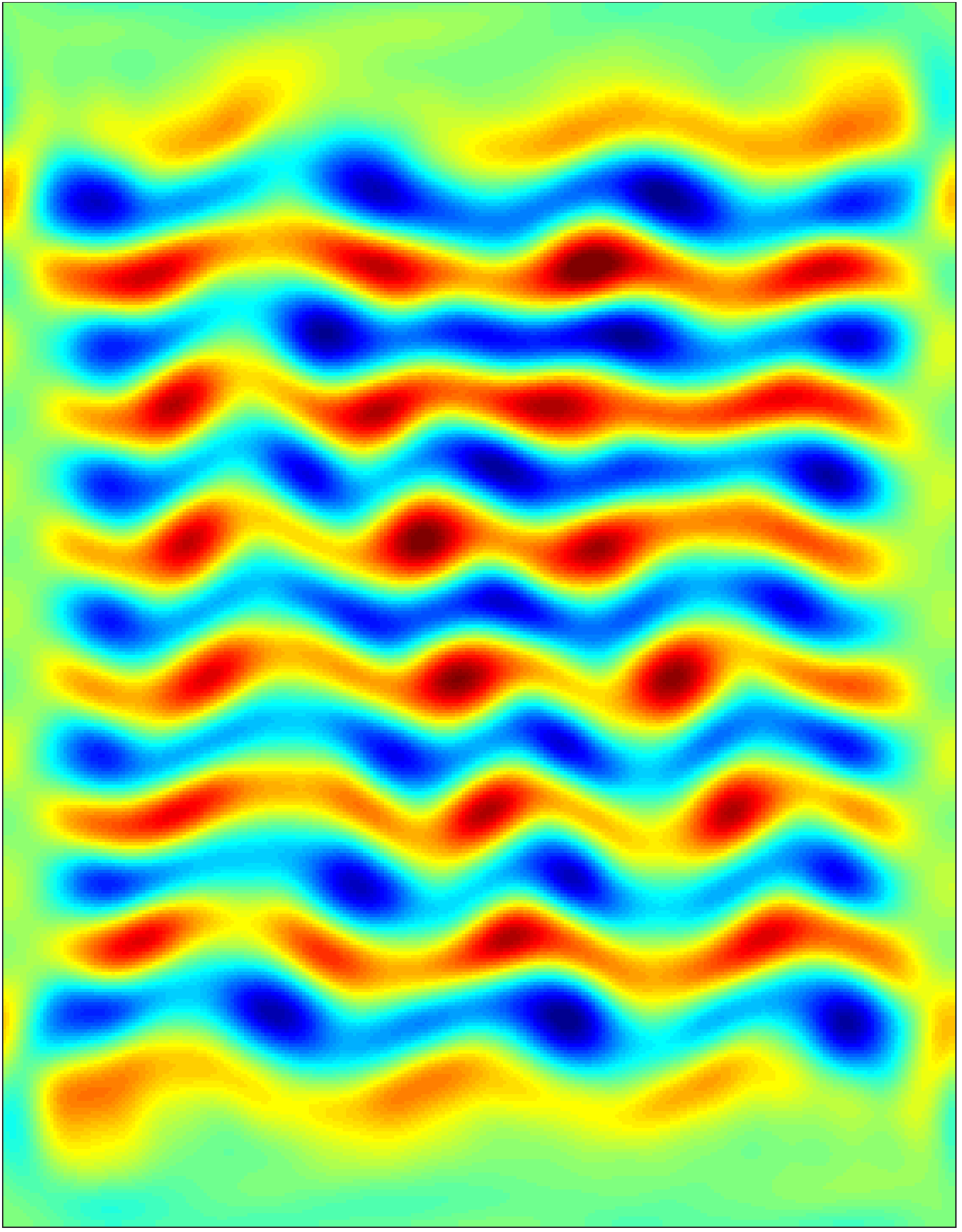}}\hspace{1.75mm} 
\subfloat[]{\includegraphics[height=2in]{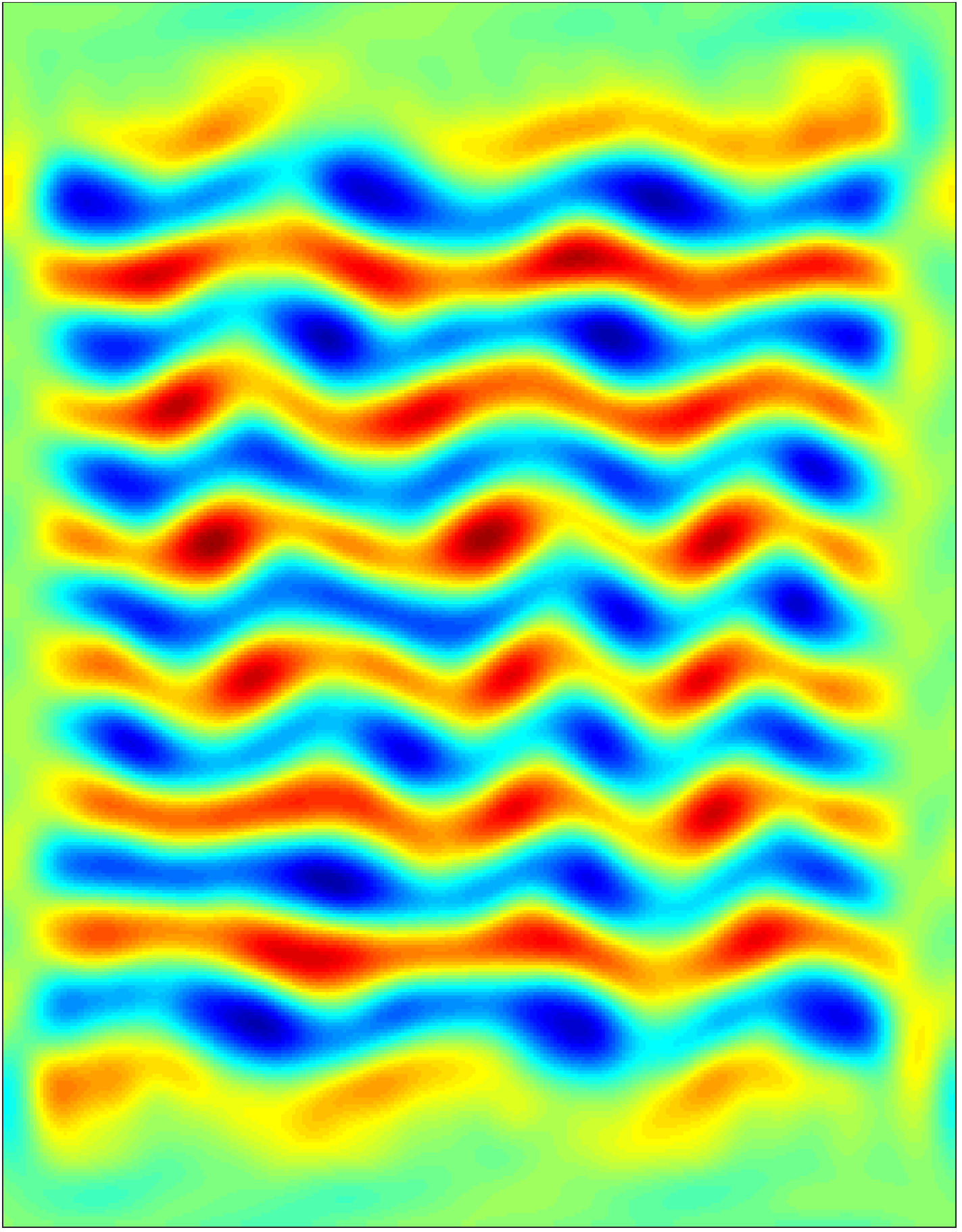}}
\caption{\label{fig:flow_fields_q1dm} {Comparison of flow fields far from E03. States on 1D submanifold trajectories (a) ${\bf u}_+(t'_m)$, marked using red diamond in \reffig{manifold_1d} and (c) ${\bf u}_-(t'_m)$ ({red square}).  Turbulent flow fields  ${\bf u}(t_m)$ from (b) simulation (black diamond on $\textrm{T}_4$) and (d) experiment (green square on $\textrm{T}_5$).}}
\end{figure}

Both turbulent trajectories shown in \reffig{manifold_1d} follow the 1D submanifold quite closely in the full state space, as the plot of the distance $D_1(t)$ shown in \reffig{1d_follow} illustrates.
In particular, the numerical trajectory follows ${\bf u}_{+}(t')$ with $D_1\leq 0.3$ along the entire interval $-0.5\tau_c\lesssim t\lesssim 3.5\tau_c$ shown in \reffig{manifold_1d}. 
Note that, once again, we set $t=0$ at the instant when $s(t)$ achieves a minimum for each of the turbulent trajectories visiting the neighborhood of E03.
The experimental trajectory follows the 1D submanifold over a shorter interval $-0.5\tau_c\lesssim t\lesssim 2.5\tau_c$, after which it diverges from the submanifold as indicated by the value of $D_1$ exceeding our empiric threshold of 0.6.
Notice that ${\bf u}(t)$ initially follows ${\bf u}_{+}(t')$, but at $t\approx\tau_c$ it switches and starts following ${\bf u}_-(t')$. 
At that point ${\bf u}_+(t')$ and ${\bf u}_-(t')$ are themselves quite close.

To illustrate how well the 1D submanifold reproduces turbulent dynamics far away from E03, we compare in \reffig{flow_fields_q1dm}(a) and (b) the flow field at the instant $t'_m$ marked using the red diamond on ${\bf u}_+(t'_m)$ in \reffig{manifold_1d} with the nearest point (black diamond) on the turbulent trajectory ${\bf u}(t_m)$ from simulation. Similar comparison between ${\bf u}_-(t'_m)$ and  ${\bf u}(t_m)$ in experiment (red and green squares in \reffig{manifold_1d}) is shown in \reffig{flow_fields_q1dm}(c) and (d).
The instant $t_m$ for each turbulent trajectory corresponds to the smaller of the distances
\begin{align}\label{eq:eq_1dtoturb2}
D^\pm_1 = \min_t\frac{\|{\bf u}(t)-{\bf u}_\pm(t'_m)\|}{\|{\bf u}(t)\|}
\end{align}
from ${\bf u}(t)$ to ${\bf u}_\pm(t'_m)$.

These flow fields associated with ${\bf u}_\pm(t'_m)$ in the physical space are significantly different from that associated with E03 (cf. \reffig{ecs_q1dm}), confirming that ${\bf u}_\pm(t'_m)$ are far away from E03 in state space. 
This can be quantified more directly in terms of the distance to ${\bf u}_\pm(t'_m)$ along the submanifold, defined as an arc length in state space
\begin{align}
D_m=\frac{1}{\|{\bf u}_0\|}\int_{-\infty}^{t'_m} \left\|\frac{d{\bf u}_\pm(t')}{dt'}\right\|{dt'}\approx \frac{1}{\tau_c}\int_{-\infty}^{t'_m} s(t')dt',
\end{align}
where ${\bf u}_0={\bf u}_\pm(-\infty)$ is the equilibrium.
The corresponding distances $D_m\approx 2.16$ for ${\bf u}_+(t'_m)$ (red diamond) and $D_m\approx 2.3$ for ${\bf u}_-(t'_m)$ (red square) are substantially larger than the empiric limit of $D_0\approx 0.6$ for closeness in state space.
Hence we can conclude that the submanifold guides the evolution of these turbulent trajectories over large distances in the full state space, {not just in its low-dimensional projection shown in figure \ref{fig:manifold_1d}.}

\begin{figure}[!t]
\centering
\includegraphics[width=3.3in]{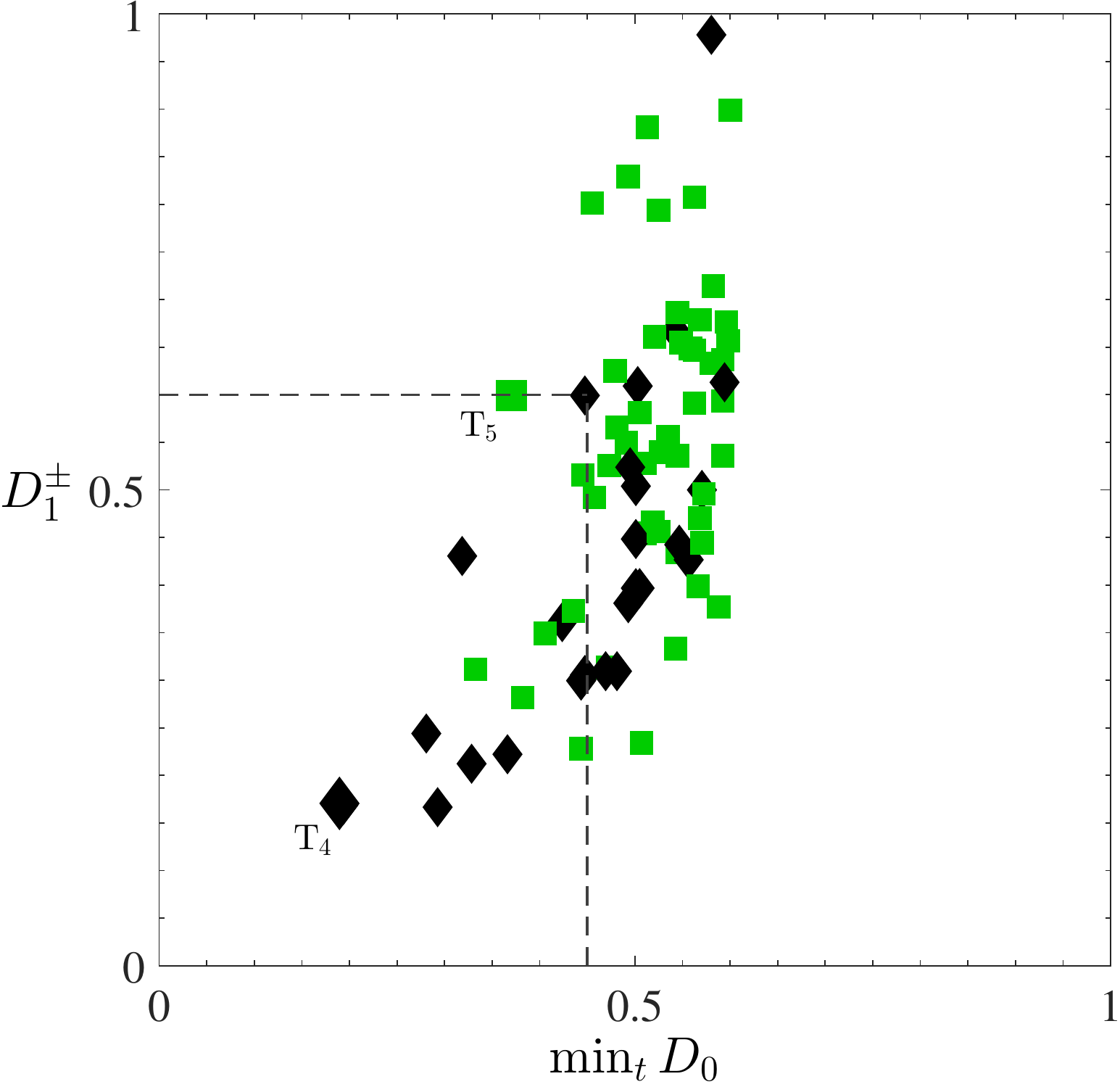}
\caption{\label{fig:q1dm_statistics} Distances from turbulent trajectories to the 1D submanifold. Axes $\min_t D_0(t)$ and $D^\pm_1$ are the minimum distances to E03 and ${\bf u}_\pm(t'_m)$, respectively. {Diamonds ($D_1^+$) and squares ($D_1^-$)} {indicate closeness} to ${\bf u}_+$ and ${\bf u}_-$, respectively.}
\end{figure}

In the discussion so far, we have demonstrated the dynamical role of the 1D submanifold using a pair of turbulent trajectories, one in simulation and the other in experiment, that approach E03 the closest.
In fact, we found that {\it all} turbulent trajectories that come sufficiently close to this equilibrium always depart following its 1D submanifold.  
In the numerical simulations, we identified about 75 distinct instances when turbulent trajectories came within a distance $D_0\leq 0.6$ of E03.  
To check whether each trajectory follows the 1D submanifold after it leaves the neighborhood of E03, we compute the {distances $D^\pm_1$} to the reference points ${\bf u}_+(t'_m)$ {(red diamond)} and ${\bf u}_-(t'_m)$ {(red square)}.
Comparing {these distances,} we can identify whether a turbulent trajectory follows ${\bf u}_+$ or ${\bf u}_-$.

Figure \ref{fig:q1dm_statistics} shows {$D^\pm_1$} for each of the 75 trajectories versus $\min_t D_0(t)$.
Following the notations of figures \ref{fig:manifold_1d} and \ref{fig:1d_follow}, we use {black} diamonds (green squares) to indicate that ${\bf u}(t)$ follows ${\bf u}_+$ (${\bf u}_-$). 
For reference, $D_1^+$ ($D_1^-$) corresponding to the trajectory  $\textrm{T}_4$ ($\textrm{T}_5$) is labeled in  \reffig{q1dm_statistics}.
All trajectories that approach E03 within a distance $D_0 \leq 0.45$ follow the 1D submanifold quite closely, with $D^\pm_1 \leq 0.6$. 
Hence, although the 1D submanifold is not locally attracting, it shapes the geometry of a relatively large region of state space around E03. 
Even for $D_0 \geq 0.45$, a large fraction of turbulent trajectories still follows the 1D submanifold ($D^\pm_1\leq 0.6$); however the submanifold stops being a reliable predictor of the evolution.

\section{Conclusions}\label{sec:summary}

A vast majority of studies investigating fluid turbulence from a dynamical systems perspective have focused on the role of ECSs -- mainly unstable traveling waves and time-periodic states --  in the transition between laminar flow and turbulence, primarily using numerical simulations. 
In comparison, the dynamical role of other types of invariant sets (e.g., equilibria and stable/unstable manifolds associated with different ECSs) in turbulent evolution has received very little attention.
Our combined experimental and numerical investigation of a canonical two-dimensional flow suggests that these invariant sets 
are important themselves.

Specifically, we found that unstable equilibria are responsible for the frequently observed slowdowns in the evolution of a weakly turbulent flow, making {their presence} felt far outside of their linear neighborhoods.
On the other hand, no dynamically relevant time-periodic solutions with periods of up to 30 correlation times were found, despite a thorough and systematic search. 
By computing how closely turbulent trajectories approach each equilibrium, we have also determined that 
at least 27 of the 31 equilibria that we have computed are dynamically relevant, marking the first time a key piece of the dynamical systems approach has been directly validated in an experimental setting.

While equilibria themselves, being point-like objects in the state space, cannot shape the geometry of the chaotic set underpinning {fluid turbulence}, the associated stable and unstable manifolds do.
We have demonstrated that {unstable} manifolds guide the evolution of turbulent {flows} that happen to pass through {rather large neighborhoods} of the corresponding equilibria, which makes {these manifolds} important building blocks in the {deterministic, geometrical description of turbulence}.

In particular, {\it unstable} manifolds associated with the dynamically dominant equilibria are relatively low-dimensional, which substantially constrains the shape of trajectories in their vicinity.
The evolution can be constrained even further for unstable manifolds with a leading eigenvalue which is well-separated from the rest.
As an example, we have shown that several turbulent trajectories entering the neighborhood of an equilibrium with a seven-dimensional unstable manifold depart following a one-dimensional submanifold associated with the leading eigenvector.

The dynamical role of {\it stable} manifolds of equilibria has not been discussed in any detail in the present study.
They do appear to play an important role, however.
A recent numerical study of channel flow has shown that the hairpin vortices -- perhaps the most recognizable example of coherent structures in wall-bounded turbulent fluid flows -- 
actually arise due to transient amplification of small, but finite-amplitude, disturbances in the stable manifold associated with a traveling wave solution (relative equilibrium) \cite{farano_2018}. 
This intriguing result suggests that a more rigorous exploration of the role of stable manifolds is necessary for a better understanding of state space geometry.

The shape of both stable and unstable manifolds plays a key role in generating the chaotic dynamics underpinning turbulence. 
Heteroclinic tangles of stable and unstable manifolds of different ECSs are responsible for the stretching and folding of state space volumes that is an essential mechanism of chaos.
On the other hand, intersections of stable and unstable manifolds of different ECSs define the heteroclinic connections, which connect neighborhoods of different ECSs and are expected to guide and constrain the evolution of turbulent flow as it moves between these neighborhoods. 
While the role of heteroclinic connections in fluid turbulence has also received little attention, they could potentially provide as much insight into the inner workings of turbulence as ECSs do.
The present study is but a first step in computing and understanding the dynamical role of these underappreciated invariant {sets}.

\begin{acknowledgments}
This work is supported by grants from the National Science Foundation (CMMI-1234436, DMS-1125302, CMMI-1725587) and Defense Advanced Research Projects Agency (HR0011-16-2-0033).
\end{acknowledgments}

\appendix
\section{State Space Projections}\label{sec:projections}

\begin{figure}[!htbp]
\centering
\subfloat[]{\includegraphics[width=1.02in]{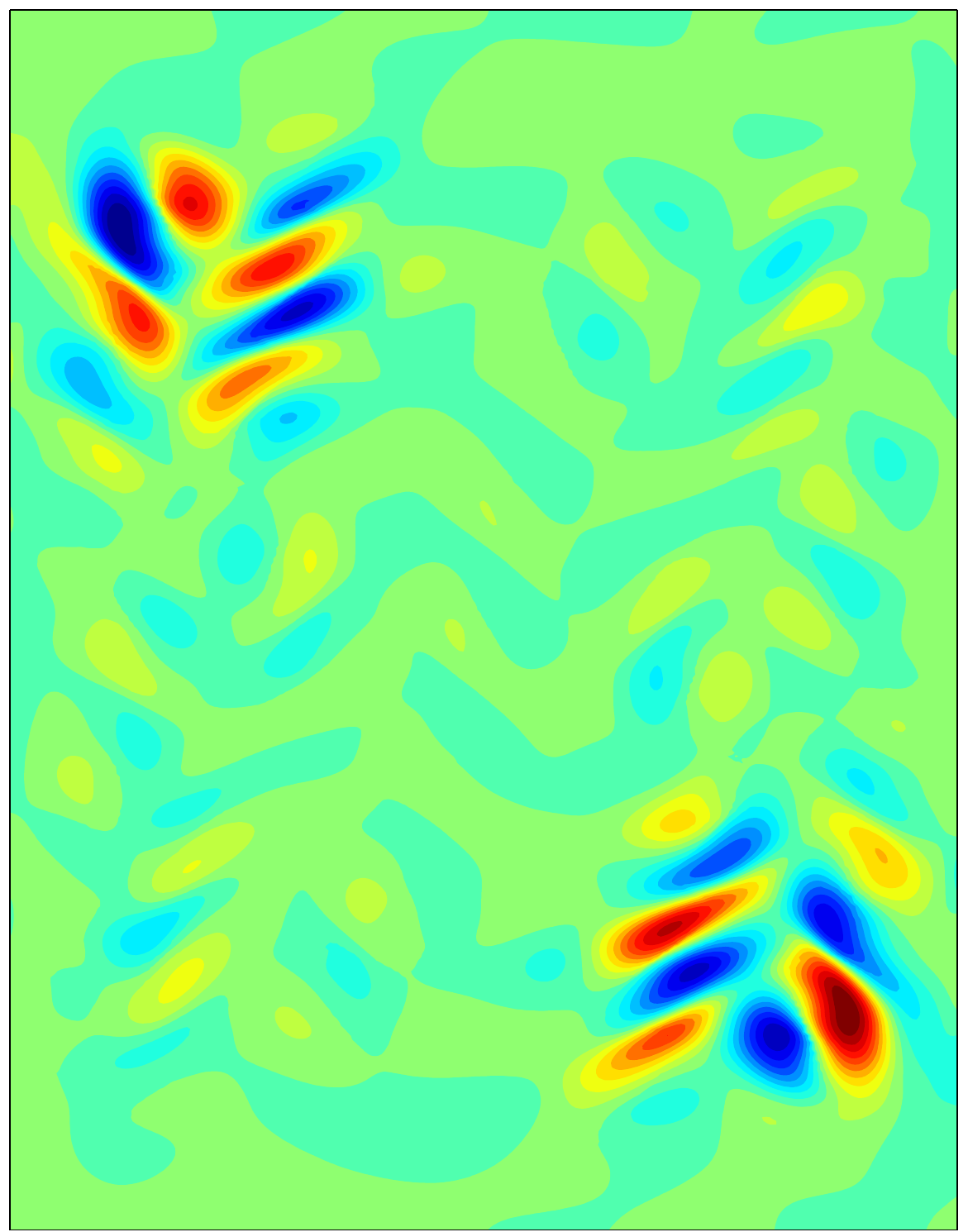}} \,
\subfloat[]{\includegraphics[width=1.02in]{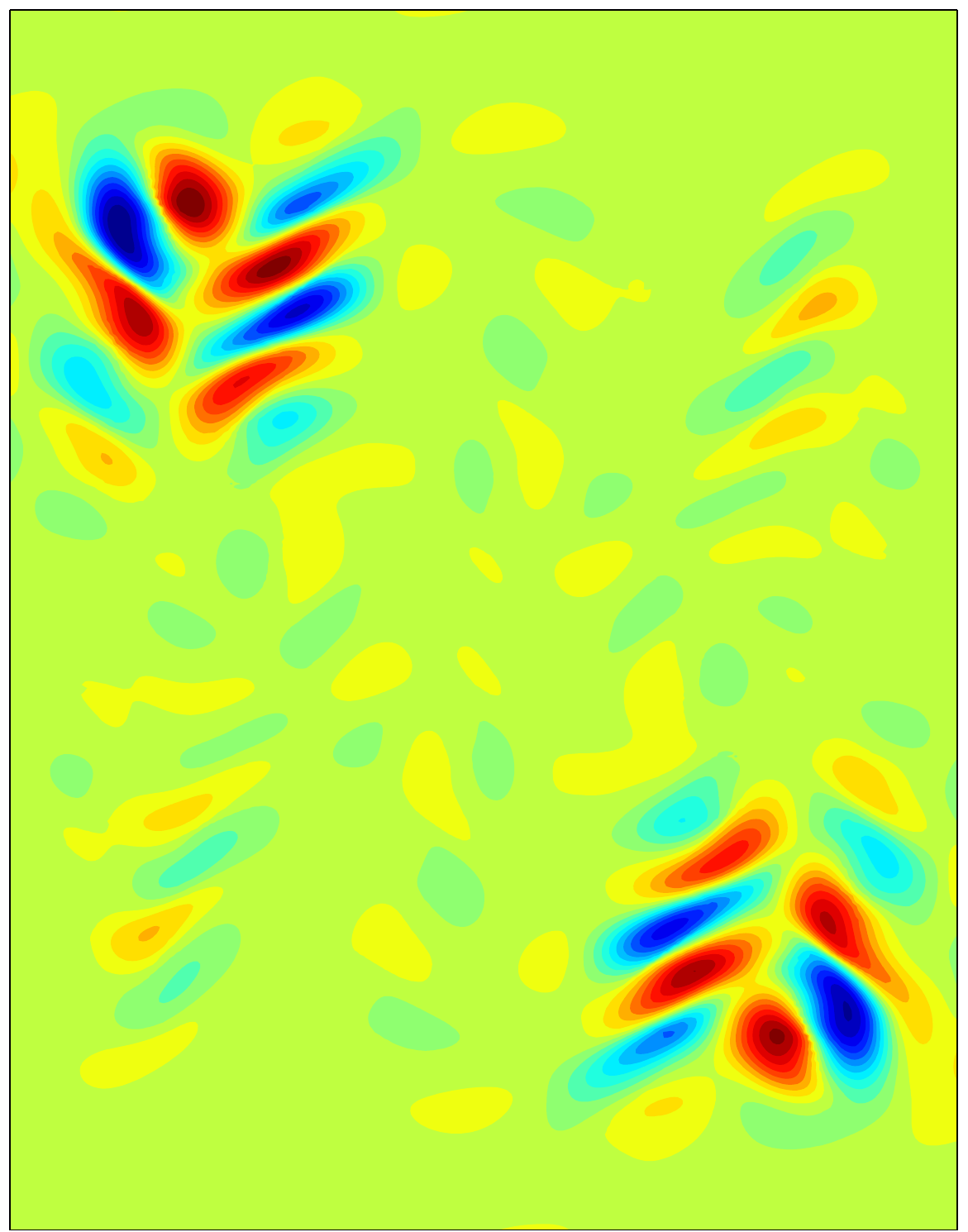}} \,
\subfloat[]{\includegraphics[width=1.02in]{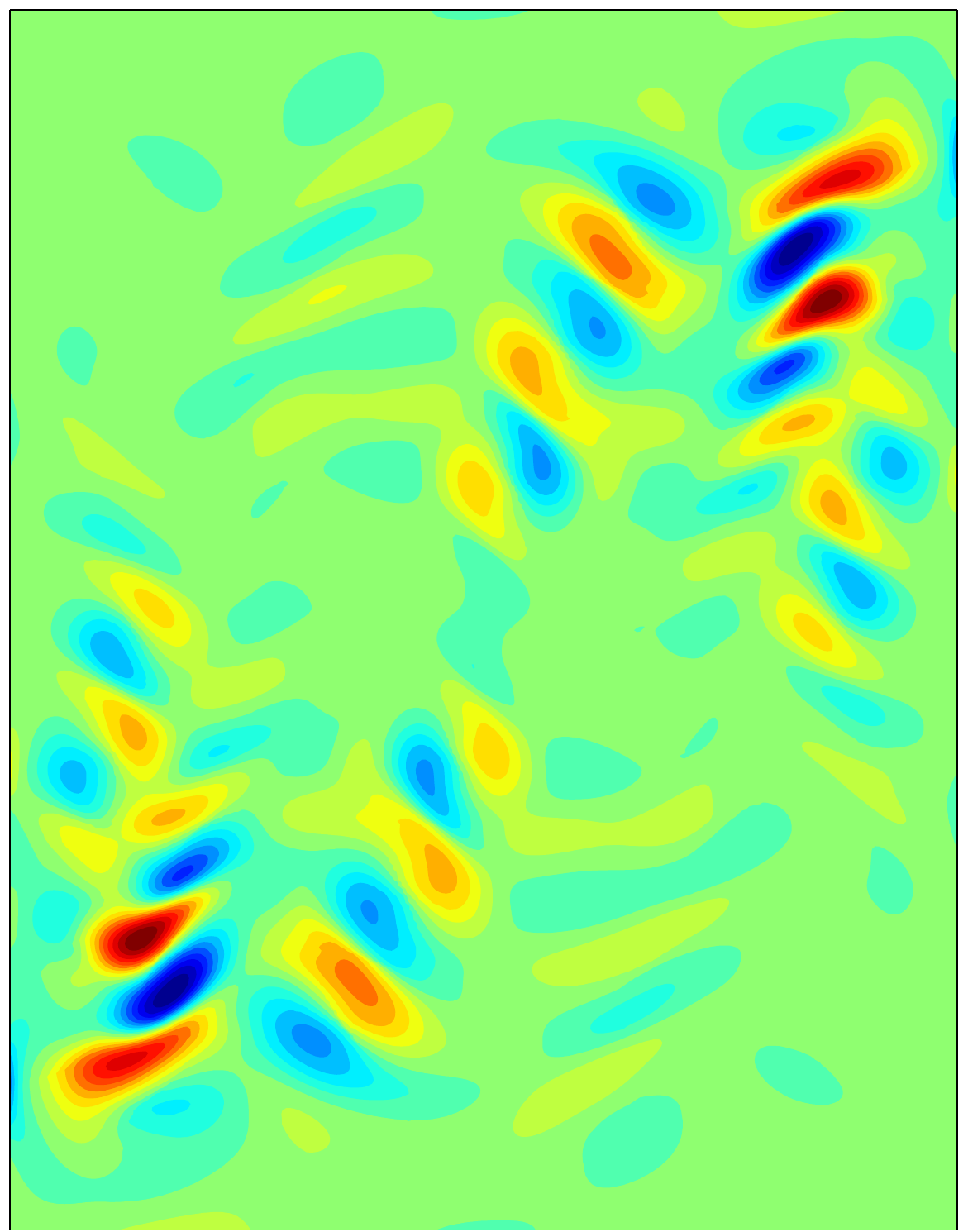}}
\caption{\label{fig:eigenvectors_2dm} Eigenvectors of E01 used to construct the 3D projection of the state space shown in \reffig{manifold_2d}: (a) $\uv{e}_1$, (b) $\uv{e}_2$, (c) $\uv{e}_5$.} 
\end{figure}
\begin{figure}[!htbp]
\centering
\subfloat[]{\includegraphics[width=1.02in]{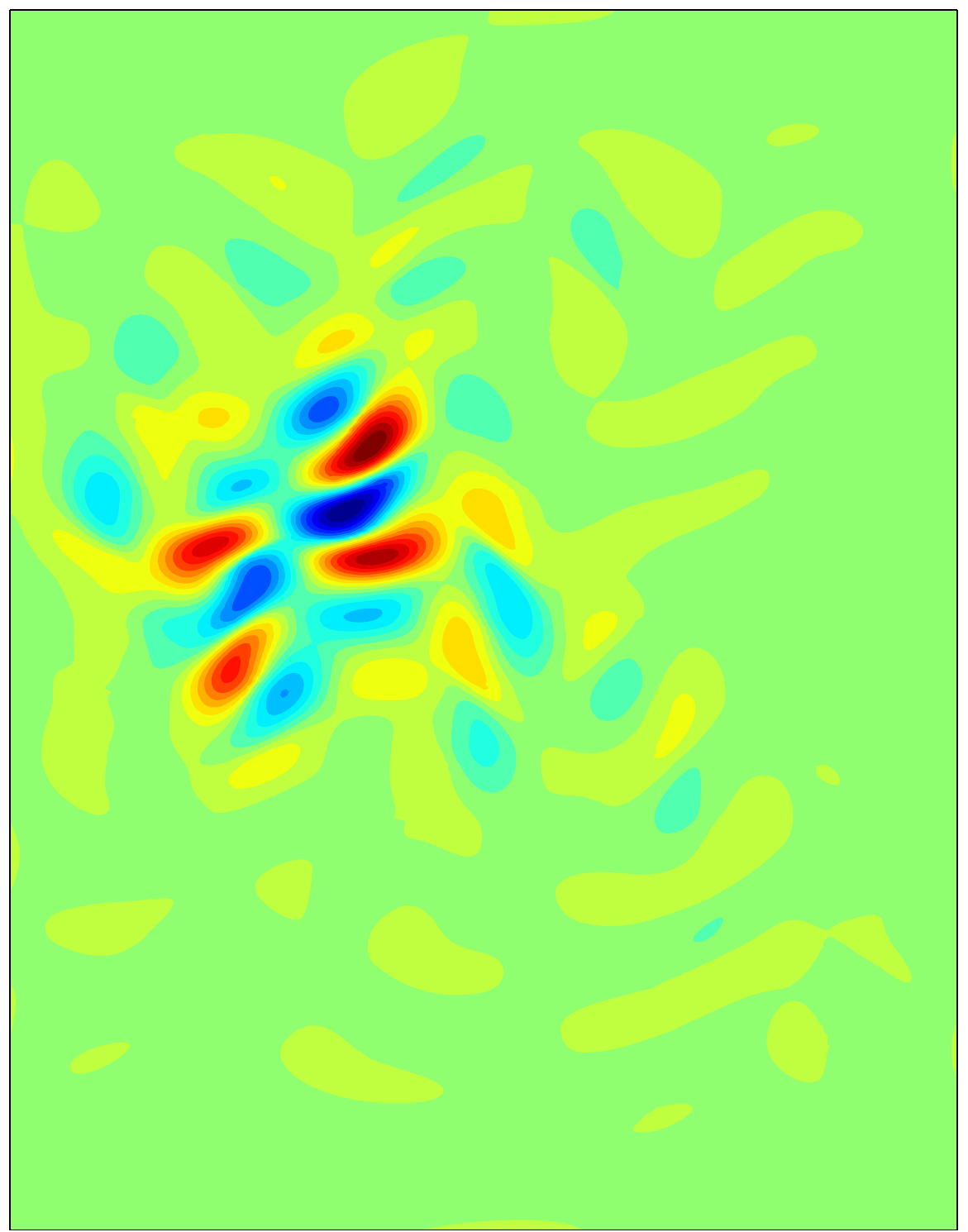}} \,
\subfloat[]{\includegraphics[width=1.02in]{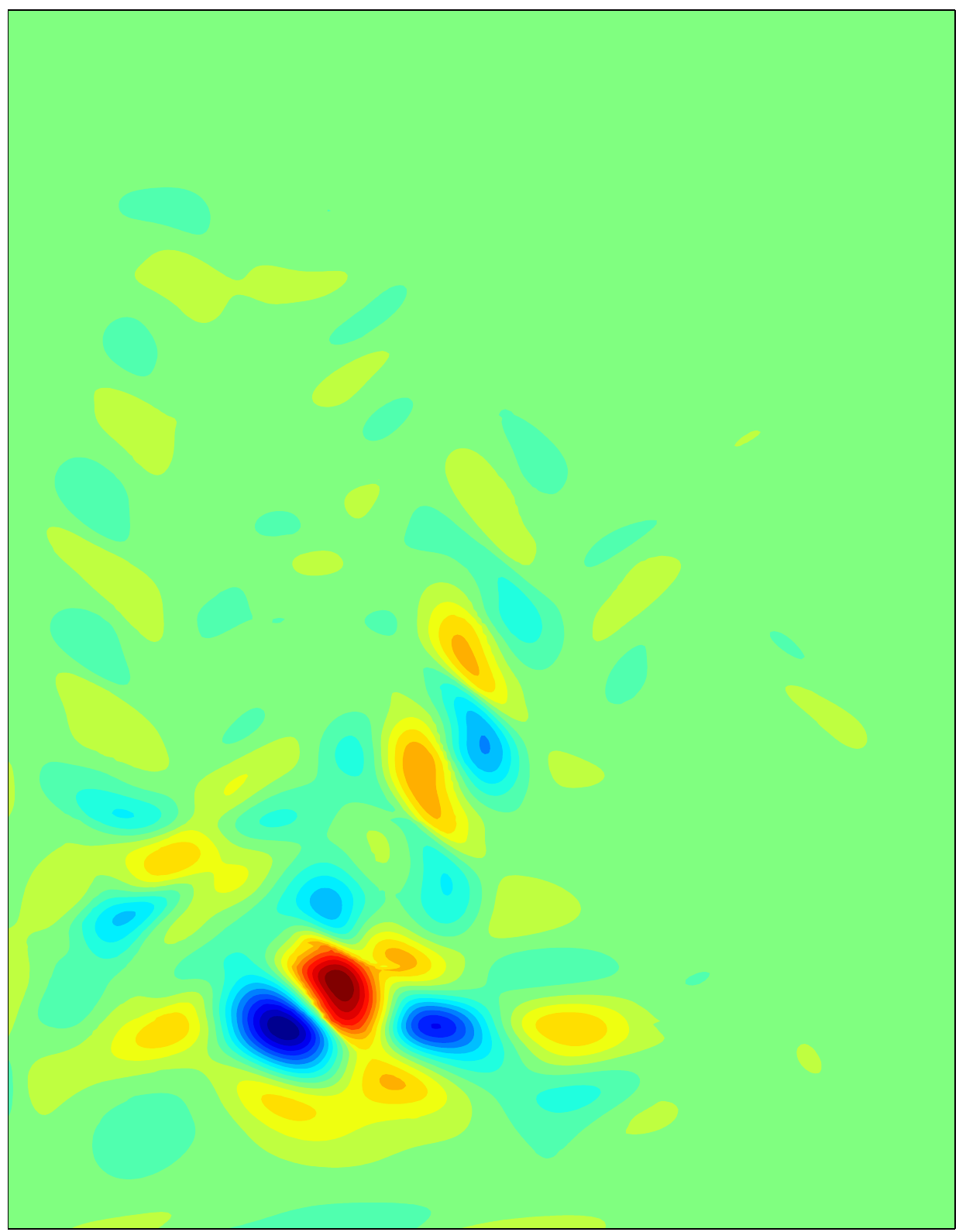}} \,
\subfloat[]{\includegraphics[width=1.02in]{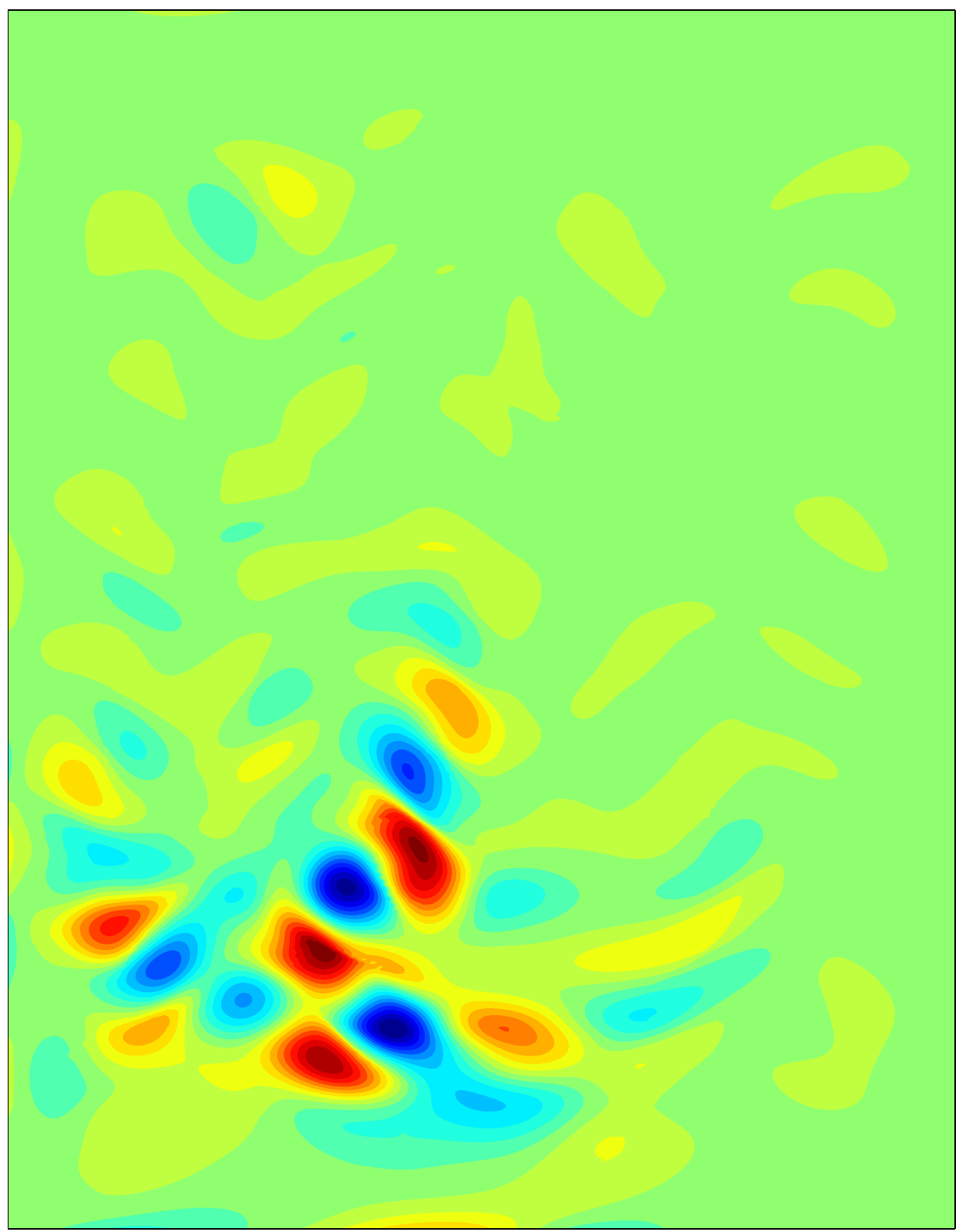}}
\caption{\label{fig:eigenvectors_1dm} Eigenvectors of E03 used to construct the 3D projection of the state space shown in \reffig{ecs_q1dm}: (a) $\uv{e}_1$, (b) $(\uv{e}_6+\uv{e}_7)/2$, (c) $(\uv{e}_6-\uv{e}_7)/(2i)$.} 
\end{figure}

In this appendix, we provide details on constructing state space visualizations in figures \ref{fig:manifold_2d} and \ref{fig:manifold_1d}. 
Near an equilibrium ${\bf u}_0$, we express the turbulent and manifold trajectories in state space as a linear combination of the eigenvectors $\uv{e}_k$ of the equilibrium:
\begin{equation}\label{eq:eq_lincomb}
{\bf u}(t) = {\bf u}_0+\sum_k a_k\uv{e}_k
\end{equation}
Typically, eigenvectors $\uv{e}_k$  are not mutually orthogonal, i.e., $\uv{e}_k\cdot\uv{e}_l \neq \delta_{k,l}$, where $\delta_{k,l}$ is the Kronecker delta. 
Hence, coordinates $a_k$ are given by using the scalar product
\begin{equation}
a_k = \uv{e}_k^\dagger\cdot({\bf u}(t) -{\bf u}_0)
\end{equation}
{with adjoint eigenvectors $\uv{e}_k^\dagger$ such that $\uv{e}_k^\dagger\cdot\uv{e}_l=\delta_{k,l}$. }
To project the state space onto a subspace spanned by any three eigenvectors $\uv{e}_k,\uv{e}_l,\uv{e}_m$,  we construct orthonormal vectors $\uv{e}^\prime_k = T_{kl}\uv{e}_l$. Here, $T_{kl}$ is computed using the orthonormality condition $\uv{e}^\prime_k\cdot\uv{e}^\prime_l = \delta_{k,l}$. The coordinates
\begin{equation}
c_k =T_{kl} a_l
\end{equation}
along  vectors $\uv{e}^\prime_k$ are then {used} to generate the projection figures.

The choice of the three eigenvectors is guided by their dynamical relevance as well as the amplitudes $a_k$ of nearby turbulent trajectories.
In particular, the 2D unstable manifold of E01 shown in \reffig{manifold_2d}
is locally tangent to the plane spanned by $\uv{e}_1$ and $\uv{e}_2$, so these two eigenvectors are a natural choice. 
For the third direction, we chose the stable eigenvector $\uv{e}_5$, since the manifold trajectories far away from E01 {tend to} have large components along $\uv{e}_5$. This proved useful in showcasing the curvature of the unstable manifold.
The shapes of $\uv{e}_1$, $\uv{e}_2$, and $\uv{e}_5$ in the physical space are shown in \reffig{eigenvectors_2dm}.

In the neighborhood of E03 (cf. \reffig{manifold_1d}), we project the state space onto a basis spanned by the leading unstable {eigenvector} $\uv{e}_1$ and the eigenvectors $\uv{e}_6$, $\uv{e}_7$ associated with the complex conjugate eigenvalue pair $\lambda_6=\lambda_7^*$. 
Like in the 2D manifold example, eigenvectors $\uv{e}_6$, $\uv{e}_7$ were chosen because they well represent the curvature of the 1D submanifold away from E03.
The shapes of $\uv{e}_1$ and the real and complex parts of $\uv{e}_6, \uv{e}_7$ in physical space are shown in \reffig{eigenvectors_1dm}.  

\section{Temporal Autocorrelation}\label{sec:sec_auto_corr}
\begin{figure}[!b]
\centering
{\includegraphics[width=3.3in]{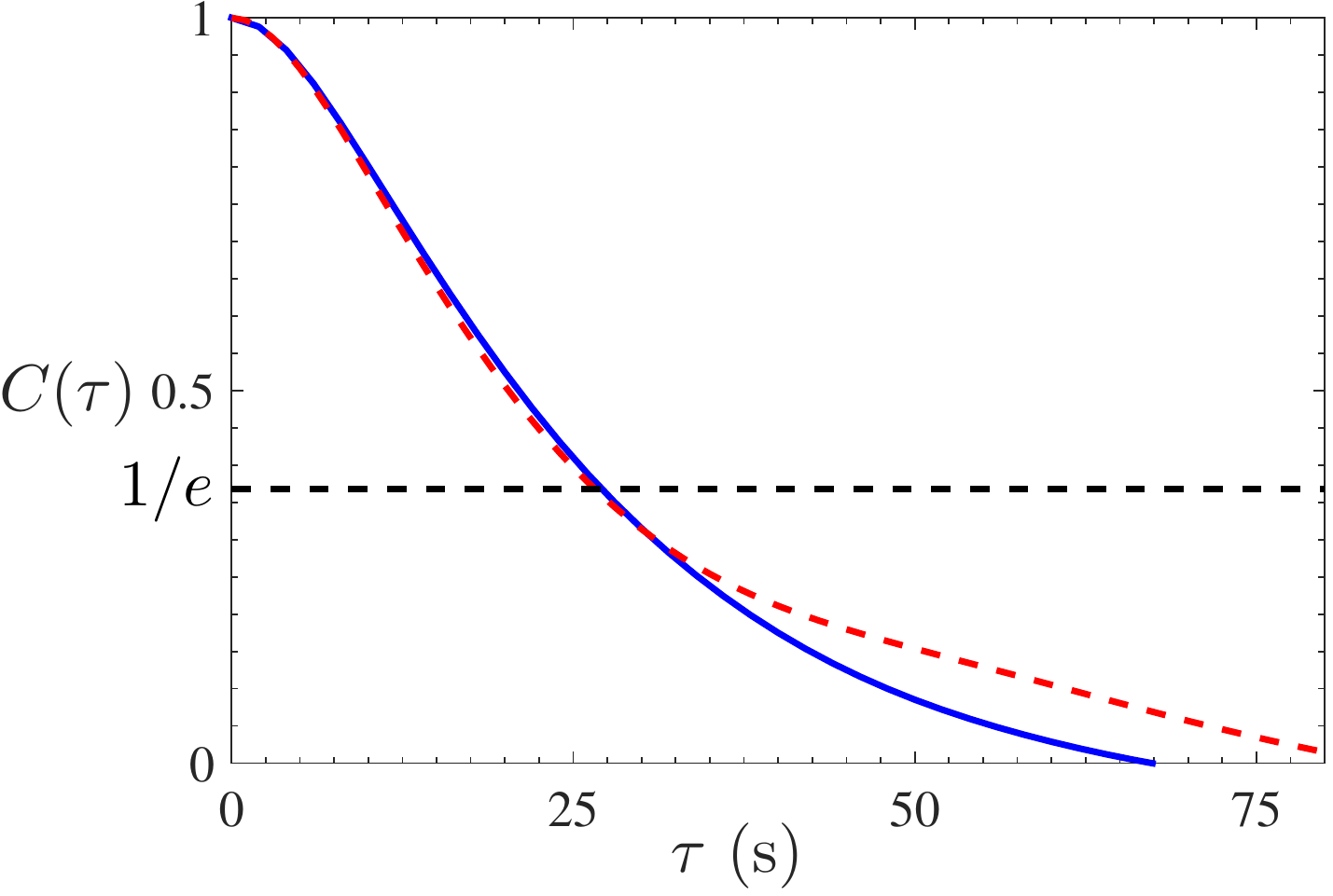}}
\caption{\label{fig:auto_correlation} Temporal auto-correlation of the velocity field at $Re = 22.5$. {The {solid (dashed)} curve corresponds to simulation (experiment).}
}
\end{figure}

The estimate for how long it typically takes for a turbulent flow field to change significantly in the course of its evolution can be obtained from the temporal auto-correlation of the velocity field:
\begin{equation}\label{eq:auto_corr}
C(\tau) = \frac{\langle \Delta{\bf u}(t)\cdot \Delta{\bf u}(t+\tau) \rangle_t}{\langle \Delta{\bf u}(t)\cdot \Delta{\bf u}(t) \rangle_t},
\end{equation}
where $\langle\cdot\rangle_t$ indicates temporal average, $\Delta {\bf u}(t) = {\bf u}(t) - \langle{\bf u}(t)\rangle_t$, and the scalar product of two fields ${\bf u}$ and ${\bf v}$ is defined as
\begin{equation}\label{eq:dot_product}
{\bf u}\cdot{\bf v}=\sum_{i=x,y}\int_\Omega u_i({\bf x})v_i({\bf x})d^2{\bf x},
\end{equation}
where $\Omega$ is the flow domain. 

Figure \ref{fig:auto_correlation} shows a plot of the normalized temporal auto-correlation as a function of $\tau$ at $Re = 22.5$. The normalization criterion $C(0) =1$, takes into account that a flow field at every instant is identical (and hence perfectly correlated) to itself. The correlation time $\tau_c$ {can be defined as the} smallest root of $C(\tau) = 1/e$, denoted by the black dashed line in \reffig{auto_correlation}. For the simulation as well as the experiment the correlation time $\tau_c \approx 12.5$ ($27\pm 1$ s in dimensional units).

\end{document}